\documentclass{article}
\usepackage[margin=1in]{geometry}

\usepackage{amssymb}

\usepackage{amsmath}
\DeclareMathOperator*{\cell}{cell}
\newcommand{\unif}{\mathcal{U}}

\usepackage[T1]{fontenc}
\usepackage{graphicx}
\usepackage{natbib}
\usepackage{ragged2e}
\usepackage{filecontents}

\usepackage{hyperref}
\hypersetup{
colorlinks=true,
linkcolor=blue,
citecolor=blue
}

\usepackage[capitalize]{cleveref}

\usepackage{algorithm}
\usepackage{algpseudocode}
\Crefname{ALC@unique}{Line}{Lines}

\usepackage[table]{xcolor}
\usepackage{booktabs}
\usepackage{multirow}
\usepackage{array}
\usepackage[caption=false]{subfig}
\usepackage{siunitx}
\usepackage{listings}
\lstdefinestyle{lststyle}{
    commentstyle=\color{cyan},
    keywordstyle=\color{orange},
    basicstyle=\ttfamily\footnotesize,
    breakatwhitespace=false,
    breaklines=true,
    captionpos=b,
    keepspaces=false,
    showspaces=false,
    showstringspaces=false,
    showtabs=false,
    tabsize=2,
    frame=tb,
    xleftmargin=.08\textwidth,
    xrightmargin=.08\textwidth,
    linewidth=0.34\linewidth,
    identifierstyle=\color{black},
    language=Fortran
}

\crefname{lstlisting}{Lst.}{Lsts.}
\Crefname{lstlisting}{Listing}{Listings}

\title{On the performance of two-sided MPI, MPI-3 RMA and SHMEM in a Lagrangian
particle cluster algorithm}

\author{Matthias Frey\thanks{Mathematical Institute, University of St Andrews, KY16 9SS, UK},
Douglas Shanks\thanks{Hewlett Packard Enterprise, London, EC2V 7HR, UK},
Steven B\"oing\thanks{School of Earth and Environment, University of Leeds, LS2 9JT, UK},
Rui F. G. Ap\'{o}stolo\thanks{EPCC, The University of Edinburgh, EH8 9BT, UK}}

\begin{document}

\maketitle

\begin{abstract}
In this paper, we compare the parallel performance of three distributed-memory communication models for a cluster algorithm based on a nearest neighbour search algorithm for $N$-body simulations. The nearest neighbour is defined by the Euclidean distance in three-dimensional space. The resulting directed nearest neighbour graphs that are used to define the clusters are pruned in an iterative procedure where we use either point-to-point message passing interface (MPI), MPI-3 remote memory access (RMA), or SHMEM communication. The original algorithm has been developed and implemented as part of the elliptical parcel-in-cell (EPIC) method targeting geophysical fluid flows. The parallel scalability of the algorithm is discussed by means of an artificial and a standard fluid dynamics test case. Performance measurements were carried out on three different computing systems with InfiniBand FDR, Hewlett Packard Enterprise (HPE) Slingshot 10 or HPE Slingshot 200 interconnect.
\end{abstract}

\subsection*{\normalsize\bfseries Keywords}
distributed-memory parallelism, cluster algorithm, nearest neighbour search, N-body simulation, one-sided communication, remote memory access (RMA), MPI, SHMEM

\section{Introduction}
The identification of nearby or nearest neighbours is fundamental to problems in
statistical learning (e.g.\ classification and regression) and in force calculations of $N$-body simulations (e.g.\ molecular dynamics, smoothed-particle hydrodynamics, and particle-in-cell). While the former considers closeness in terms of data similarity, the latter uses the spatial distance between two finite-size or point particles as a measure.
Previous work on finding $k$-nearest neighbours includes, for example, \citet{Friedman1975, Fukunaga1975, Yuval1975, Yuval1976, KamgarParsi1985}. A review of various techniques for finding near-neighbours within a fixed radius in a $d$-dimensional Euclidean space is provided by \citet{Bentley1975}. A more recently published work on the fixed radius neighbour search by \citet{Chen2024} uses sorting to reduce the search space and therefore improve the performance of the algorithm. This paper focusses on the $k = 1$ nearest neighbour search (NNS) in 3-dimensional Euclidean space associated with particle-particle interactions in $N$-body simulations. For more general NNS algorithms, we refer the interested reader to the articles mentioned above and references therein.\\
A well-known application of the more general case with multiple neighbours is the Verlet list (or neighbour list) \citep{Verlet1967}, used in molecular dynamics to reduce the computational complexity for the evaluation of short-ranged non-bonded interactions, often fitted by a Lennard-Jones potential. For each particle, the method stores all neighbouring particles within a cut-off radius in a list or array data structure and evaluates the resulting force on a particle based solely on the stored neighbours. A similar approach  is the cell-linked list (or cell list) method \citep{Quentrec1973}. This method divides the domain into isotropic grid cells and evaluates the short-range forces on a particle by iterating through all the particles inside that same cell and its neighbouring cells. Hence, this method creates a list of particles per cell rather than per particle as for the Verlet list. Both cell lists and the Verlet list are used for other applications, such as fluid dynamics simulated with smoothed particle hydrodynamics \citep{Dominguez2011}. In the calculation of gravitational or electrostatic forces as they appear in cosmology or plasma simulations, respectively, the domain partitioning, as carried out with the fast multipole method (FMM) or the Barnes-Hut algorithm, naturally leads to tree data structures. These tree-based methods enable the efficient evaluation of the short- and long-range force contributions. Here, we only focus on interactions between the nearest neighbours. In particle-mesh methods (e.g.\ particle/parcel-in-cell), on the other hand, the domain is already partitioned into usually isotropic grid cells. All particles within a grid cell can therefore be obtained by simple integer arithmetic.\\
Here, we report the parallel performance of a cluster algorithm based on a NNS that is used to merge nearby parcels (finite-sized particles) as part of the elliptical parcel-in-cell (EPIC) \citep{Frey2022, Frey2023} method. The EPIC method is based on the idea of using deformable Lagrangian parcels of elliptical shape in order to better capture the dynamics of turbulent fluid flows. Elongated parcels are split into two in order to maintain the effective subgrid scale resolution of the flow, and to avoid numerical problems that arise from parcels with large aspect ratio. To reduce computational cost, very small parcels obtained through successive splitting are merged with their closest neighbour. The NNS algorithm involves two
stages: the first step constructs the nearest neighbour graphs (NNGs), where only small parcels point to their nearest neighbour, which can either be a small or big parcel. In the second step, the NNGs are split in an iterative procedure to avoid parcel chains that would result in elongated ellipsoids. While we use standard point-to-point communication during the NNG construction step, the pruning stage uses either point-to-point or one-sided (remote memory access) communication with MPI-3 RMA or SHMEM. The various versions of the cluster algorithm were developed outside the original code base of EPIC
and all source codes of this study are openly available as indicated in the section on code availability.\\
A related work in the context of machine learning by \citet{Lin2015} also uses a hybrid MPI and SHMEM approach to improve the performance of a $k$-nearest neighbour algorithm.
Previous studies evaluating the performance of MPI RMA and SHMEM routines on various computing systems include work by \citet{Luecke2004} and \citet{Negoita2017}. While the former compared MPI-2 RMA with SHMEM on a SGI
Origin 2000 and a Cray T3E-600, \citet{Negoita2017} assessed the performance and scalability of MPI-3 RMA and SHMEM routines for message sizes of 8 bytes, 10 KB and 1 MB on a Cray XC30 with Aries interconnect. They concluded that SHMEM routines generally outperformed two-sided and one-sided MPI-3 routines on their system. Their test about ``accessing distant messages'', where the root process receives single messages from other processes, seems closest to our application. Table 1 in
\citep{Negoita2017} shows that for small message sizes, i.e.\ 8 bytes, SHMEM get/put are considerably faster than MPI-3 RMA get/put. However, with increasing message size, their timings converge. Since our application exchanges 4-byte messages, we might therefore expect SHMEM to perform noticeably better than MPI-3 RMA. \citet{Li2014} report scaling results of MPI-3 RMA for a breadth-first search using the Graph500 benchmark. Overall, their study shows a speed-up in the graph traversal time of about 2 for 4,096 cores with MPI-3 RMA compared to previous MPI implementations with MPI-2 RMA or two-sided MPI. In their study, they also demonstrate that a put operation (\verb|MPI_Put|) together with the new MPI-3 RMA feature \verb|MPI_Win_flush| improves performance over MPI-2 RMA with \verb|MPI_Win_fence|. \citet{Gerstenberger2013, Gerstenberger2018} developed the MPI-3 RMA library \textit{foMPI} (fast one-sided MPI) for Cray Gemini (XK5, XE6) and Aries (XC30) where they use bufferless protocols (i.e.\ no remote buffering) over remote direct memory access (RDMA) networks. While for inter-node communication the RDMA operations are performed with DMAPP (Distributed Memory Application), the intra-node communication uses the Linux kernel module XPMEM. Note that for this study we only compare library-based frameworks for distributed parallel computing and therefore do not consider Coarray Fortran (CAF) which is part of the Fortran language. For comparisons between CAF and MPI, we refer the interested reader to \citep{Shterenlikht2019, Garain2015}. Also, \citet{Rouson2017} present performance results for an atmospheric model enhanced with CAF based on OpenCoarrays \citep{Fanfarillo2014} using MPI or OpenSHMEM as backend.\\
In order to assess the parallel performance of the different communication models as part of our cluster algorithm, we carried out scaling studies on three different computing systems having either an InfiniBand FDR, HPE Slingshot 10 or HPE Slingshot 200 interconnect. The parallel scalability is discussed using artificial data and data generated from a standard fluid dynamics test case.\\
This paper is organised as follows: in \cref{sec:nns} we first summarise the nearest neighbour cluster algorithm and explain its parallelisation in detail. This is followed by benchmark results in \cref{sec:benchmarks}. That section also includes latency and bandwidth results for the different interconnects. Final remarks are provided in \cref{sec:conclusions}.

\section{Nearest neighbour cluster algorithm}
\label{sec:nns}

\begin{figure}[!htp]
    \centering
    \includegraphics{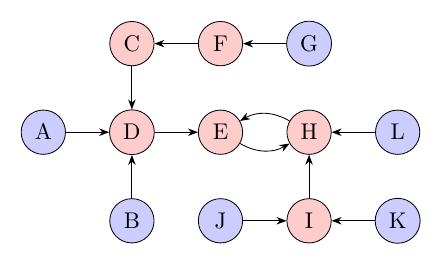}
    \caption{An example of an unweighted directed graph (DG). We call all nodes without an incoming edge \textit{leaf vertices}. This DG has leaf vertices A, B, G, J, K and L.}
    \label{fig:dg}
\end{figure}

The nearest neighbour cluster algorithm described here was developed as part of the EPIC method \citep{Frey2022, Frey2023}. The algorithm consists of two major steps: the directed graph (DG) construction and the DG pruning. The DG construction step identifies the nearest neighbour based on the Euclidean distance. For this purpose, each object (an ellipsoid in our case) is assigned to its nearest grid point on a Cartesian node-centred mesh. It should be noted that each grid cell contains many parcels, usually of the order of 20. This domain discretisation reduces the operation count of the nearest neighbour search from $N^2$ to $8N_{\cell}$, where $N$ is the total number of objects and $N_{\cell}$ represents an average number of objects per grid box in 3D space. For each object, we then search for its nearest neighbour over all the surrounding grid cells and establish a unidirectional link. If two objects point to each other, we call this a dual link. The terminology we use here follows that in previous publications on EPIC. A more extensive discussion of some of the properties of nearest neighbour graphs is also given in \citealt{Eppstein1997}. This includes generalisation from 1 to $k$ nearest neighbours. For a set of points that are connected in the graph, there can only be one such dual link, since each parcel only points to one other parcel. Longer cycles are excluded (this could only happen with 3 or more parcels at the exact same distance from each other, which is extremely unlikely). After the search operation is complete, we are left with unweighted directed graphs. An example of a DG is shown in \cref{fig:dg}. A directed graph $G = (V, E)$, defined as a set of vertices $V$ and a set of edges $E$ obtained through the aforementioned procedure, has the following properties:
\begin{itemize}
    \item The outdegree of each vertex $v\in V$ is $\deg^{+}(v) \le 1$. A parcel which does not need merging has $\deg^{+}(v) = 0$.
    \item The indegree of a vertex $v\in V$ is $\deg^{-}(v) \ge 0$.
\end{itemize}
We denote an edge pointing from vertex $v_1$ to vertex $v_2$ by $e_{12} = (v_1, v_2)$.
A vertex $v\in V$ with indegree $\deg^{-}(v) = 0$ is called a leaf. In \cref{fig:dg} we colour all leaf vertices, A, B, G, J, K and L, in blue. We denote the collection of leaf vertices by $\mathbb{L} := \{v\in V|\deg^{-}(v) = 0\}$. We further define the set of vertices available for merging by
$\mathbb{A} := \mathbb{L} \cup \{v\in V \vert \deg^{-}(v) > 0 \wedge \forall e_j = (v_j, v) \in E, v_j\in \mathbb{L}\}$, i.e. all vertices that are either a leaf or have only incoming links from leaf vertices. In terms of the graph in \cref{fig:dg}, this condition is satisfied by the vertices F and I as well as the leaf vertices mentioned before. A subgraph $G_s \subseteq G$ which only consists of vertices that are in $\mathbb{A}$ can be merged.

\begin{figure}[!htp]
    \centering
    \subfloat[Iteration 1 of stage 1: Remove the edges from F to C and I to H.]{
         \includegraphics[width=0.31\textwidth]{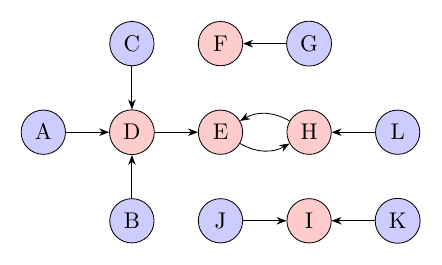}
         \label{fig:dg_resolve_iter_1}
     }
     \hfill
     \subfloat[Iteration 2 of stage 1: Remove the edge from D to E.]{
         \includegraphics[width=0.31\textwidth]{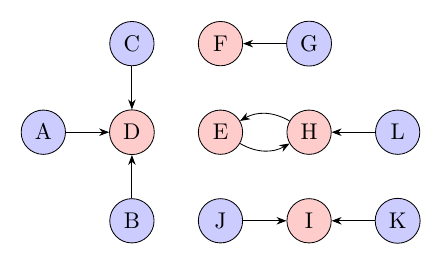}
         \label{fig:dg_resolve_iter_2}
     }
     \hfill
     \subfloat[Stage 2: Remove the edge from H to E.]{
         \includegraphics[width=0.31\textwidth]{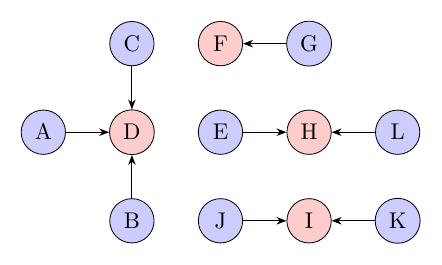}
         \label{fig:dg_resolve_iter_3}
     }
    \caption{Directed graph pruning step for the graph illustrated in \cref{fig:dg}. The algorithm consists of two stages. In the first stage, an iterative procedure performs two iterations illustrated in (a) and (b). The second stage eliminates all dual links as shown in (c). After these stages we are left with four smaller subgraphs.}
    \label{fig:dg_resolve}
\end{figure}

After the DG construction, the directed graphs are split into subgraphs $G_s$ in order to avoid large cluster chains. This two-stage process ensures that after both stages all vertices are contained in $\mathbb{A}$ such that they can be merged. The first stage is an iterative procedure where in each iteration connections between non-leaf nodes are deleted. However, this stage excludes dual links, i.e. connections between two vertices that point to each other. For the directed graph in \cref{fig:dg}, the algorithm performs two iterations as shown in \cref{fig:dg_resolve}. First, the edges from vertex F to C and I to H are removed (cf. \cref{fig:dg_resolve_iter_1}). The former edge removal results in $\deg^{-}(C) = 0$, i.e. C is a leaf vertex. In the second iteration the edge from D to E is removed (cf. \cref{fig:dg_resolve_iter_2}). After these two iterations, the original cluster is split into four smaller subgraphs. However, the subgraph consisting of vertices E, H and L features a dual link that must be simplified. The algorithm therefore enters the second stage. Here, the edge from H to E is broken up (cf. \cref{fig:dg_resolve_iter_3}) because $\deg^{-}(H) > 1$. For further details on the serial algorithm we refer to \citep[appendix D]{Frey2022}.

The parallel version of the nearest neighbour cluster algorithm presented here was developed with atmospheric simulations in mind, where the vertical dimension is kept serial. However, this only affects the graph construction step, which could be easily extended to a three-dimensional MPI domain decomposition. We further use periodic boundary conditions in both horizontal dimensions, but the method also works, subject to minor modifications, for impermeable boundary conditions.

In the first step of the algorithm, all objects that are supposed to be clustered with their nearest neighbour are determined. In EPIC, the criterion for ellipsoids to be merged is their volume: if it decreases enough, which can happen due to successive splitting, ellipsoids are merged. A change in the criterion would require only a minor modification. During this search process, the positions of objects that need merging and are near domain boundaries are sent to processes (or MPI ranks) owning adjacent domains, and the receiving MPI ranks append them to their own attribute containers. Objects received from other MPI ranks are referred to as \textit{remote objects}, and objects located in the subdomain owned by the same MPI rank are referred to as \textit{local objects}.
After this operation is complete, the MPI point-to-point and MPI-3 RMA version of the code have the option to create an MPI sub-communicator which only contains MPI ranks having small objects, i.e. local and/or remote. The aim of this sub-communicator is to reduce the overhead of MPI synchronisation and MPI collective communication during the DG pruning step when the distribution of small objects is spatially concentrated. Note that in \cref{sec:benchmarks} we only show the results without sub-communicator because there was no significant performance benefit.  To locally determine the nearest other object by evaluating the Euclidean distance, four contiguous arrays storing the indices of small objects \textit{isma}, the indices of the locally closest objects \textit{iclo}, the MPI ranks owning the closest object \textit{rclo} and the distances between the small object and its closest object \textit{dclo} are filled. Note that if no close object for remote objects is found locally, the value of \textit{dclo} is set to some big value, here $L_x^2 + L_y^2 + L_z^2$ where $L_x$, $L_y$ and $L_z$ are the domain lengths in the three Cartesian dimensions.\\
In order to determine the closest neighbour of domain-boundary objects globally, the information of remote objects, i.e. the values of \textit{isma}, \textit{iclo} and \textit{dclo}, is sent back to the original MPI rank owning the small object. The receiving MPI rank then updates the values of \textit{iclo}, \textit{dclo} and \textit{rclo} of the appropriate entry in \textit{isma} if a shorter distance is detected. After this step, the arrays \textit{isma}, \textit{iclo} and \textit{rclo} are properly set and can be used during the iterative procedure of the directed graph pruning.

\subsection{Directed graph (DG) pruning algorithm}
The graph pruning algorithm uses three boolean-typed arrays --- i.e. arrays which can store \textit{true}/\textit{false} values --- to keep track of leaf and available vertices as well as vertices that are already part of a final cluster. The pseudocode of the graph pruning algorithm is provided in \cref{alg:graph_resol}. Steps that include remote put/get operations, a collective communication or process synchronisation are marked in bold letters and with a right-pointing arrow ($\Rightarrow$). Note that the process synchronisation depends on the communication model. In case of the MPI point-to-point (P2P) communication, each synchronisation point denotes the exchange of remote information. After every step where a boolean entry in one of the arrays could have been altered, the local arrays must be updated. The MPI P2P version therefore contains an additional synchronisation point at line 15 in the algorithm. For the MPI-3 RMA and SHMEM version of the code, the synchronisation is an MPI barrier and SHMEM barrier (i.e. \verb|shmem_barrier_all|), respectively. These synchronisations only ensure that all processes completed a step before entering the next phase of the algorithm.\\
The \verb|while| loop from line 1 to line 10 in \cref{alg:graph_resol} denotes the first stage of the algorithm where all connections of a directed graph are broken up until there are only dual links left. For example, the directed graph of \cref{fig:dg} is broken up into four smaller graphs as shown in \cref{fig:dg_resolve_iter_1} where the graph consisting of vertices E, H and L still has a dual link between E and H. All other graphs that only consist of single links are considered fully pruned and therefore no longer considered in the process. Since dual links are neither contained in the set of leaf vertices $\mathbb{L}$ nor in the set of available vertices $\mathbb{A}$, the exit condition of the \verb|while| loop is fulfilled as soon as only dual links remain to be broken up. The dual links are then eliminated in the second stage of the algorithm which corresponds to lines 11--16. Special care must be taken with isolated dual links. These are graphs that only consist of two vertices that point to each other. They are pruned separately because there is no information from its subgraph in order to break up one of the links. For example, the graph in \cref{fig:dg_resolve_iter_2} with the dual link between E and H has the additional connection between H and L. This latter connection uniquely defines the order of elimination, i.e. the link from H to E is deleted. With isolated dual links there is no such decisive logic, which is why each of the two links can be removed. To prevent the removal of both links if the two vertices do not belong to the same MPI rank, the MPI rank with the lower process number eliminates its connection. In the following sections, we explain the different implementations of the graph pruning algorithm.
\begin{algorithm}[!htp]
    \caption{Graph pruning}\label{alg:graph_resol}
    \algnewcommand\algorithmicnot{\textbf{not}}
    \small
\algdef{SE}[IF]{IfNot}{EndIf}[1]{\algorithmicif\ \algorithmicnot\ #1\ \algorithmicthen}{\algorithmicend\ \algorithmicif}%

    \begin{algorithmic}[1]
        \While{$\mathbb{A}\ne\emptyset \wedge \mathbb{L}\ne\emptyset$}
            \State Reset properties for candidate mergers
            $\Rightarrow$ \textit{\textbf{remote put}}

            \State$\Rightarrow$ \textit{\textbf{Process synchronisation}}

            \State Determine leaf parcels
            $\Rightarrow$ \textit{\textbf{remote put}}

            \State$\Rightarrow$ \textit{\textbf{Process synchronisation}}
            \State Filter unavailable vertices
            $\Rightarrow$ \textit{\textbf{remote put}}

            \State$\Rightarrow$ \textit{\textbf{Process synchronisation}}

            \State Identify mergers
            $\Rightarrow$ \textit{\textbf{remote get and put}}

            \State$\Rightarrow$ \textit{\textbf{Process synchronisation (MPI P2P only)}}

            \State Evaluate while-loop condition $\Rightarrow$ \textit{\textbf{MPI allreduce}}
        \EndWhile

        \State Mark non-leaf parcels as available
        $\Rightarrow$ \textit{\textbf{remote put}}

        \State$\Rightarrow$ \textit{\textbf{Process synchronisation}}

        \State Prune dual links
        (\textit{\textbf{remote get}})

        \State$\Rightarrow$ \textit{\textbf{Process synchronisation}}

        \State Prune isolated dual links
        $\Rightarrow$ \textit{\textbf{remote get}}

         \State Remove eliminated edges such that arrays are contiguous
    \end{algorithmic}
\end{algorithm}







\newpage
\subsubsection{Graph pruning algorithm with MPI-3 RMA communication}
As introduced with MPI 2.0 \citep{mpi20} and then further specified with MPI 3.0 \citep{mpi30}, one-sided communication denotes a paradigm change from the classical point-to-point communication where both sender and receiver have information about the metadata of the message. One-sided communication is well suited for situations where the receiving process (or target) does not or is not required to know the sending process (or origin). Despite this flexibility, one-sided communication comes with its own challenges, mainly the synchronisation of memory accesses to avoid race conditions. There are two synchronisation methodologies: \textit{active} and \textit{passive} target communication. While active target synchronisation was introduced with MPI-2 RMA, passive target synchronisation was added with MPI-3 RMA. In passive target communication the origin process locks and unlocks the memory location on the receiving (or target) process, which does not engage in the synchronisation. Active target communication, on the other hand, still involves the receiving process, but only the sender possesses information about the metadata. In both cases a so-called \textit{epoch} denotes the time frame in which remote memory accesses are carried out. An epoch with active target communication starts and ends with a collective call to \verb|MPI_Win_fence| on the \textit{window}, i.e.\ a dedicated memory region accessible by other processes, as shown in \cref{lst:rma_active_comm}. On the other hand, an epoch with passive target communication is encapsulated by the non-collective calls to \verb|MPI_Win_lock| and \verb|MPI_Win_unlock| on a window as specified in \cref{lst:rma_passive_comm}. Here, we make use of passive target communication because not all MPI ranks as part of the communicator may be involved in the algorithm performing RMA operations. We also use \verb|MPI_Win_flush| to ensure completion of RMA operations. Note that remote get (\verb|MPI_Get|) and remote put (\verb|MPI_Put|) operations should not be applied to the same window during a single epoch as this may result in race conditions. Instead, multiple epochs should be used. Several epochs for different windows can, however, be active at the same time, so that remote operations for different memory regions can overlap.
\begin{figure*}[!htp]

    \centering
    \subfloat[Active target communication.]{
        \centering
\begin{lstlisting}[]
! Begin RMA epoch:
call MPI_Win_fence(win, ...)
! Perform RMA operations:
call MPI_Put(win, ...)
! End RMA epoch:
call MPI_Win_fence(win, ...)
\end{lstlisting}
        \label{lst:rma_active_comm}
    }
    \hspace{15mm}
    \subfloat[Passive target communication.]{
\begin{lstlisting}[]
! Begin RMA epoch:
call MPI_Win_lock(win, ...)
! Perform RMA operations:
call MPI_Put(win, ...)
! End RMA epoch:
call MPI_Win_unlock(win, ...)
\end{lstlisting}
            \label{lst:rma_passive_comm}
    }
    \caption{Incomplete Fortran sample code to demonstrate RMA operations with either active or passive target communication on a MPI window \textit{win}. Note: Instead of \texttt{MPI\_Put}, a call to \texttt{MPI\_Get} is also possible.}
\end{figure*}

\begin{figure}[!htp]
    \centering
    \subfloat[RMA with the separate memory model.]{
        \includegraphics[width=0.42\textwidth]{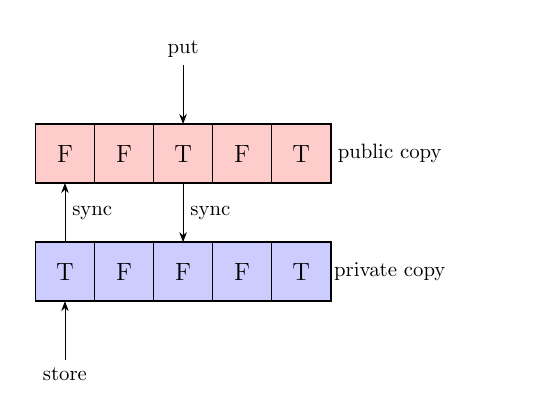}
        \label{fig:separate_memory_model}
     }
     \subfloat[RMA with the unified memory model.]{
        \includegraphics[width=0.42\textwidth]{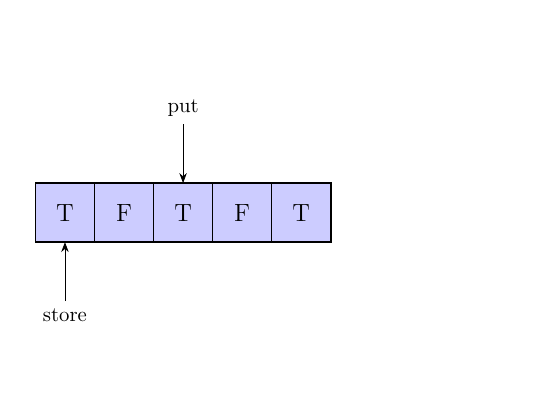}
        \label{fig:unified_memory_model}
    }
    \caption{Illustration of an RMA epoch with MPI put operations using either the separate or unified memory model for the remote memory access communication. In the unified model there is no distinction between public and private copy. In the separate memory model synchronisation calls (i.e. \texttt{MPI\_Win\_sync}) ensure memory coherence between the private and public copy. Such calls are denoted by the arrows labelled `sync'. Note that the result of sync is not reflected by the figure.}
    \label{fig:rma_memory_models}
\end{figure}

In addition to the classification of synchronisation methods, there are two types of memory models: separate and unified. The separate memory model distinguishes between private and public copy as illustrated in \cref{fig:separate_memory_model}. The private copy refers to the local memory of the owning process. The public copy is a dedicated memory region in which other processes may read (or get) and write (or put) information. A call to \verb|MPI_Win_sync| synchronises the private and public memory. In the unified memory model, on the other hand, all processes access the same memory region enabling real-time updates as depicted in \cref{fig:unified_memory_model}. Our MPI RMA implementation of the DG pruning algorithm makes use of passive target communication and uses the unified memory model.

\subsubsection{Graph pruning algorithm with MPI point-to-point communication}
\begin{figure}[!htp]
    \centering
    \subfloat[Put operations with point-to-point communication.]{
        \includegraphics[width=0.48\textwidth]{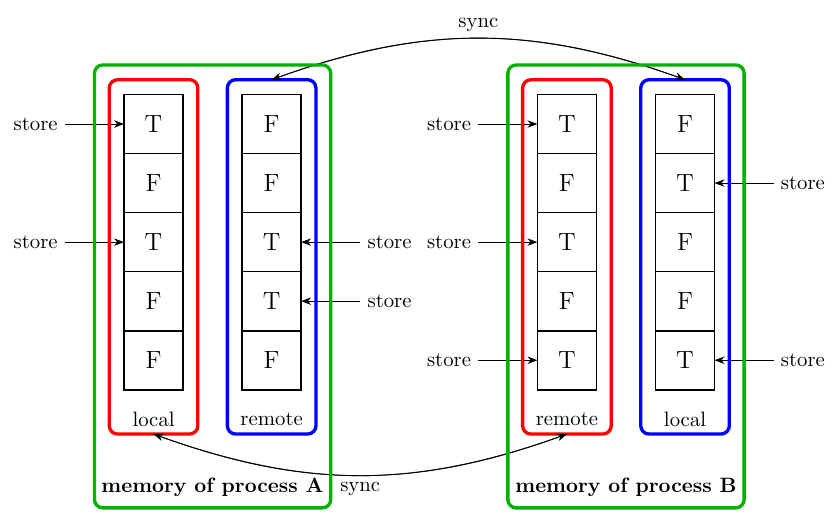}
    }
    \hfill
    \subfloat[Get operations with point-to-point communication.]{
        \includegraphics[width=0.48\textwidth]{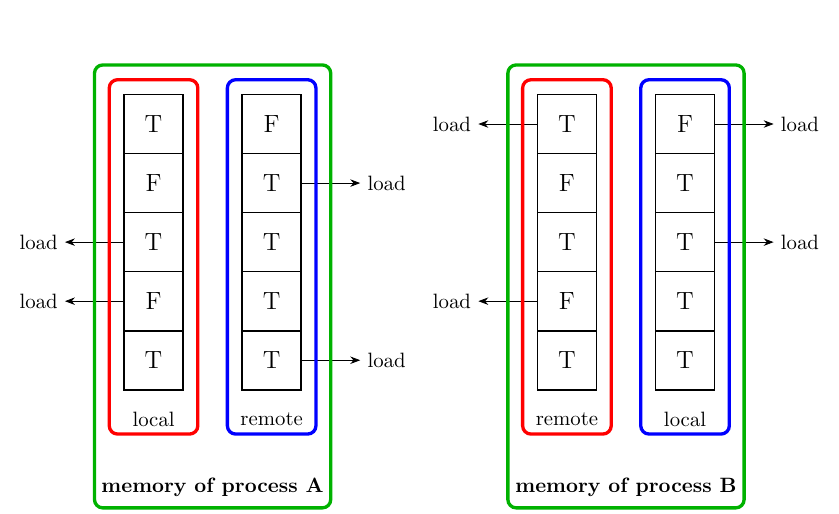}
    }
    \caption{Emulating remote memory access with point-to-point communication. Left: process A and process B write to their local and buffer memory. When an epoch ends, a synchronisation call (symbolised with the arrows labelled `sync') updates the buffer memory of process A with the local memory of process B, and vice-versa. Right: During an epoch of get operations, all accesses are performed locally, i.e.\ no communication is required.}
    \label{fig:p2p_put_get}
\end{figure}

Our version of the graph pruning with point-to-point communication basically mimics the behaviour of remote memory access with a separate memory model. For that purpose, each process stores an instance of a derived type for all its neighbouring processes. Here, neighbouring refers to the spatial location of a process based on the parallel domain decomposition. The derived type serves as a buffer for remote put and remote get operations. In an epoch of put operations, remote accesses are only performed locally by saving the new value in a buffer array of the derived type and flagging the corresponding entry as changed as illustrated in \cref{fig:p2p_put_get}. When a put epoch ends, we call a synchronisation routine that updates both the entries on the remote side and their local copies. Remote get operations are thus always performed locally without any communication. However, before performing any get operations, it must be ensured that the local data is up-to-date.

\subsubsection{Graph pruning algorithm with SHMEM communication}
SHMEM is a partitioned global address space (PGAS) model for one-sided communication introduced by Cray Inc. Over the years, several SHMEM libraries have been developed that had differences in their application programming interface (API). In order to standardise the SHMEM API, its community established the OpenSHMEM initiative \citep{Chapman2010}. Its latest API specification is OpenSHMEM 1.6.
We interface the SHMEM routines in Fortran through C bindings. For this purpose, we wrote a SHMEM module within our application similar to \citet{Linford2025} or \citet{Namashivayam2020}. Note, the latter uses Coarray Fortran (CAF) to allocate the shared memory objects. In order to exchange data of type Fortran logical (default size 4 bytes), we interfaced the SHMEM C routines \verb|shmem_put32| and \verb|shmem_get32| for remote put and remote get operations, respectively. However, if a different size is detected during the code configuration, we use \verb|shmem_putmem| and \verb|shmem_getmem| instead. In our case, the size of a Fortran logical was always 4 bytes. Similar to MPI-3 RMA, the process synchronisations in \cref{alg:graph_resol} are achieved using collective barrier calls, more precisely \verb|shmem_barrier_all|. But since the currently available OpenSHMEM team split API cannot create arbitrary subsets over all active processing elements (PEs), as is possible with MPI, the graph pruning algorithm always includes all PEs in the SHMEM version of the code. We therefore do not make use of any team API. Note, the SHMEM community refers to a process as processing element.

In the following we briefly recapitulate the developments of the relatively new OpenSHMEM team API. In 2014, \citet{Poole2014} proposed the addition of \textit{explicit active sets} to the OpenSHMEM standard. An explicit active set is a subset of PEs that can perform remote memory operations. This idea got further extended by \citet{Welch2014} to include user-defined \textit{spaces}. Instead of the name \textit{explicit active set}, \citet{Welch2014} use the \textit{teams} terminology. A \textit{team} is again a subset of PEs, and a \textit{space} is a dedicated region of symmetric memory assigned to a \textit{team} in order to perform SHMEM operations. Before the release of the OpenSHMEM 1.5 specification, \citet{Ozog2019} discussed the design and implementation of the teams API as part of the Sandia OpenSHMEM (SOS) library. They acknowledged that the idea of \textit{sets} and \textit{groups} proposed by \citet{Aderholdt2019} is more flexible in terms of the creation of PE subsets. Unfortunately, the group and set interface is not (yet) part of the OpenSHMEM specification and the currently available team split functionality as introduced with OpenSHMEM 1.5 is too limited for our application requirements.

\section{Parallel performance analysis}
\label{sec:benchmarks}
We use the three computing systems listed in \cref{tab:hardware} to analyse the parallel performance. All are equipped with a different network interconnect and processor configuration. Since we are mainly interested in the performance of the inter-node communication, the results are presented in terms of the network interconnect. Here, we use the acronym SS10, SS200 and IB for HPE Slingshot 10, HPE Slingshot 200 and InfiniBand, respectively. The toolchain on each computing system is listed in
\cref{tab:tool_chain}. The SHMEM benchmarks rely on the implementation of OpenSHMEM by Open MPI (OSHMEM) and Cray OpenSHMEMX \citep{Namashivayam2019}, a descendant of Cray SHMEM. While Cray SHMEM heavily relied on DMAPP, the underlying communication layer library SMP of Cray OpenSHMEMX differentiates between on-node data transfer using XPMEM and off-node data transfer using the OFI (OpenFabrics Initiative) transport layer. In the case of Open MPI and OSHMEM, the Unified Communication X \citep{Shamis2015,openucx-website} library is used.

\begin{table}[!htp]
    \small\sf\centering
    \caption{Computing systems and their specifications. Note that we only list the specification of standard compute nodes. For example, ARCHER2 has
    5,276 standard but also additionally 584 high memory compute nodes and 4 nodes with AMD MI210 GPUs, and Cirrus has 2 additional
    nodes with NVIDIA V100 GPUs. Also note that we use the standard CPU frequency of 2.0 GHz on ARCHER2 for all the benchmark runs.}
    \label{tab:hardware}
    \begin{center}
    \begin{tabular}{>{\raggedright}p{0.24\linewidth}
                    >{\raggedright}p{0.2\linewidth}
                    >{\raggedright}p{0.22\linewidth}
                                   p{0.21\linewidth}}
        \toprule 
        specifications          & Cirrus
                                & ARCHER2
                                & Hotlum\\
        \midrule
        computing system        & HPE SGI Apollo 8600
                                & HPE Cray EX
                                & HPE Cray EX\\
        no.\ compute nodes      & 280
                                & 5,276
                                & 1,016\\
        no.\ cores per node     & 36
                                & 128
                                & 128\\
        processor (CPU)         & Intel Xeon E5-2695 (Broadwell)
                                & AMD Zen2  EPYC 7742 (Rome)
                                & AMD Zen3  EPYC 7763 (Milan)\\
        no.\ cores per CPU      & 18
                                & 64
                                & 64\\
        CPU frequency in GHz    & 2.1
                                & 2.25
                                & 2.45\\
        memory per node	in GB   & 256
                                & 256
                                & 512\\
        network interconnect (IC)  & InfiniBand FDR
                                & HPE Slingshot 10
                                & HPE Slingshot 200 \\
        IC bandwidth in GB/s    & 54.5
                                & 100
                                & 200\\
        network topology        & tree (hypercube with cut edges)
                                & dragonfly
                                & dragonfly\\
        \bottomrule
    \end{tabular}
    \end{center}
\end{table}
\begin{table}[!htp]
    \small\sf\centering
    \caption{Toolchain on the different computing systems. While we rely on the GNU compiler collection on Cirrus, ARCHER2 and Hotlum provide the Cray Compiler Environment (CCE). Note that Open MPI includes an OpenSHMEM implementation.}
    \label{tab:tool_chain}
    \begin{center}
    \begin{tabular}{llll}
        \toprule
                        & Cirrus
                        & ARCHER2
                        & Hotlum \\
        \midrule
         compiler suite & GNU 10.2.0
                        & CCE 16.0.1
                        & CCE 18.0.1\\
         MPI library    & Open MPI 4.1.6
                        & Cray MPICH 8.1.27
                        & Cray MPICH 8.1.31\\
         SHMEM library  &
                        & Cray OpenSHMEMX 11.6.1
                        & Cray OpenSHMEMX 11.7.3 \\
        \bottomrule
    \end{tabular}
    \end{center}
\end{table}

\subsection{Latency and bandwidth tests}\label{sec:osu_benchmarks}
Before studying the parallel performance of the parcel cluster algorithm, we analyse the latency and bandwidth of the MPI point-to-point (P2P), MPI-3 RMA and SHMEM routines for on-node and off-node data transfers with the OSU (Ohio State University) Micro-Benchmarks version 7.5 \citep{OSU}. While the off-node data transfer (or inter-node communication) measures the performance over the HPE Slingshot 10 (SS10), HPE Slingshot 200 (SS200) and InfiniBand (IB) interconnects, the on-node data transfer (or intra-node communication), which we ensure to occur between two cores within a single NUMA (non-uniform memory access) domain, measures the data exchange using the shared-memory region of the processor.

\begin{figure}[!htp]
    \centering
    \includegraphics[width=1.0\textwidth]{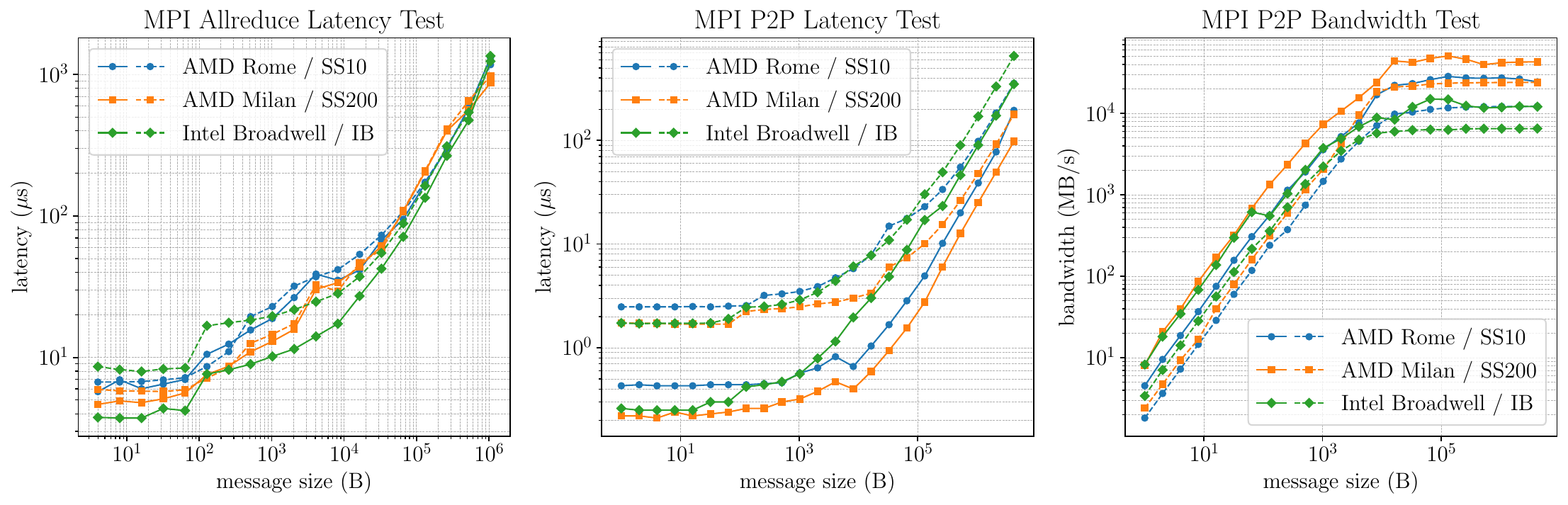}
    \caption{On-node (solid lines) and off-node (dashed lines) data transfer results of the OSU micro benchmark testing the latency and bandwidth of MPI point-to-point (P2P) communication. The left panel also shows the latency of the blocking collective MPI allreduce. The inter-node networks are HPE Slingshot 10 (SS10), HPE Slingshot 200 (SS200) and InfiniBand (IB).}
    \label{fig:osu_mpi_p2p}
\end{figure}

\begin{figure}[!htp]
    \centering
    \includegraphics[width=1.0\textwidth]{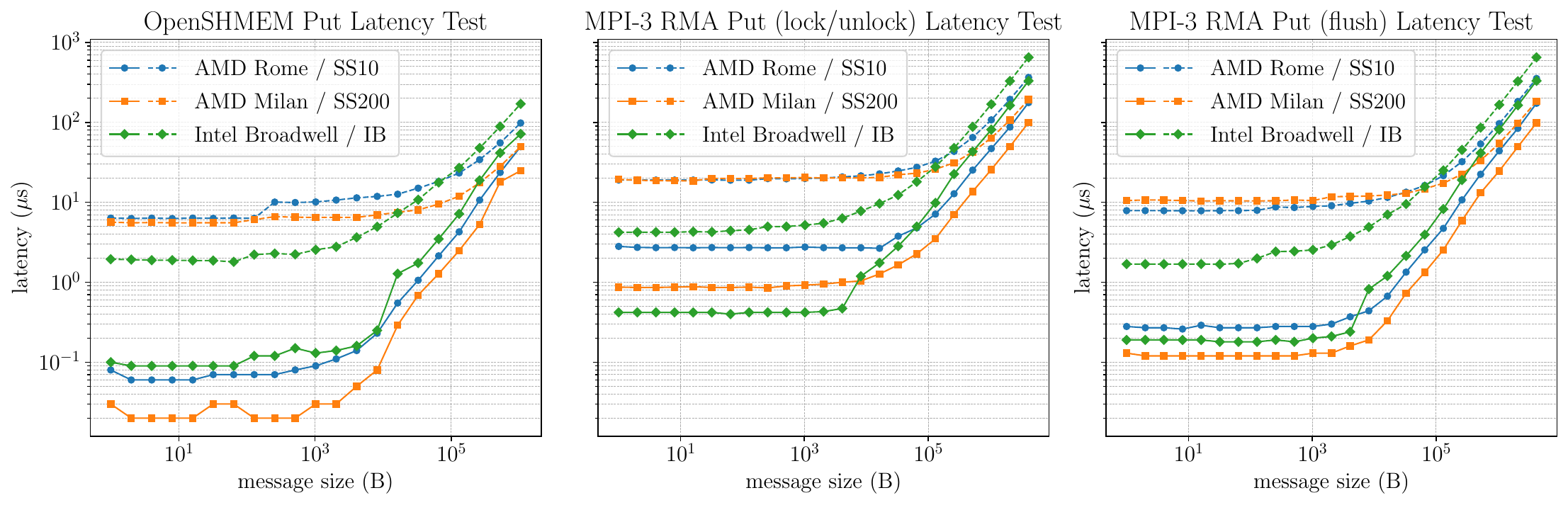}
    \caption{On-node (solid lines) and off-node (dashed lines) data transfer results of the OSU micro benchmark testing the latency of remote put operations. The inter-node networks are HPE Slingshot 10 (SS10), HPE Slingshot 200 (SS200) and InfiniBand (IB).}
    \label{fig:osu_latency_put}
\end{figure}

\begin{figure}[!htp]
    \centering
    \includegraphics[width=1.0\textwidth]{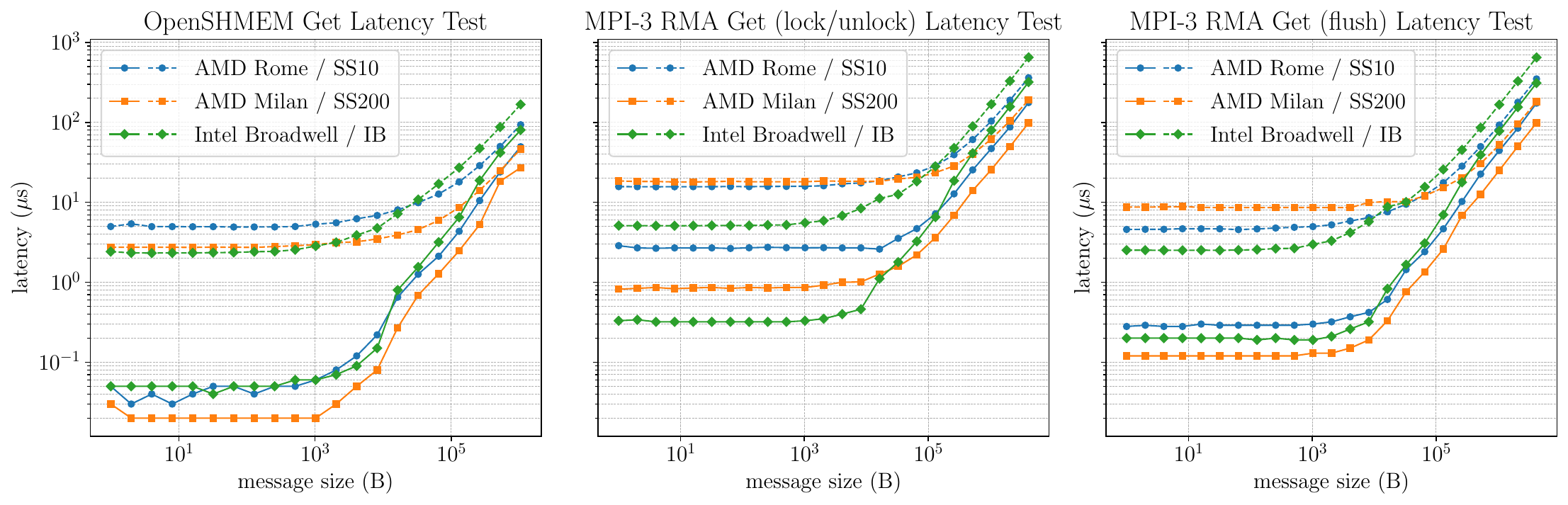}
    \caption{On-node (solid lines) and off-node (dashed lines) data transfer results of the OSU micro benchmark testing the latency of remote get operations. The inter-node networks are HPE Slingshot 10 (SS10), HPE Slingshot 200 (SS200) and InfiniBand (IB).}
    \label{fig:osu_latency_get}
\end{figure}

\begin{figure}[!htp]
    \centering
    \includegraphics[width=1.0\textwidth]{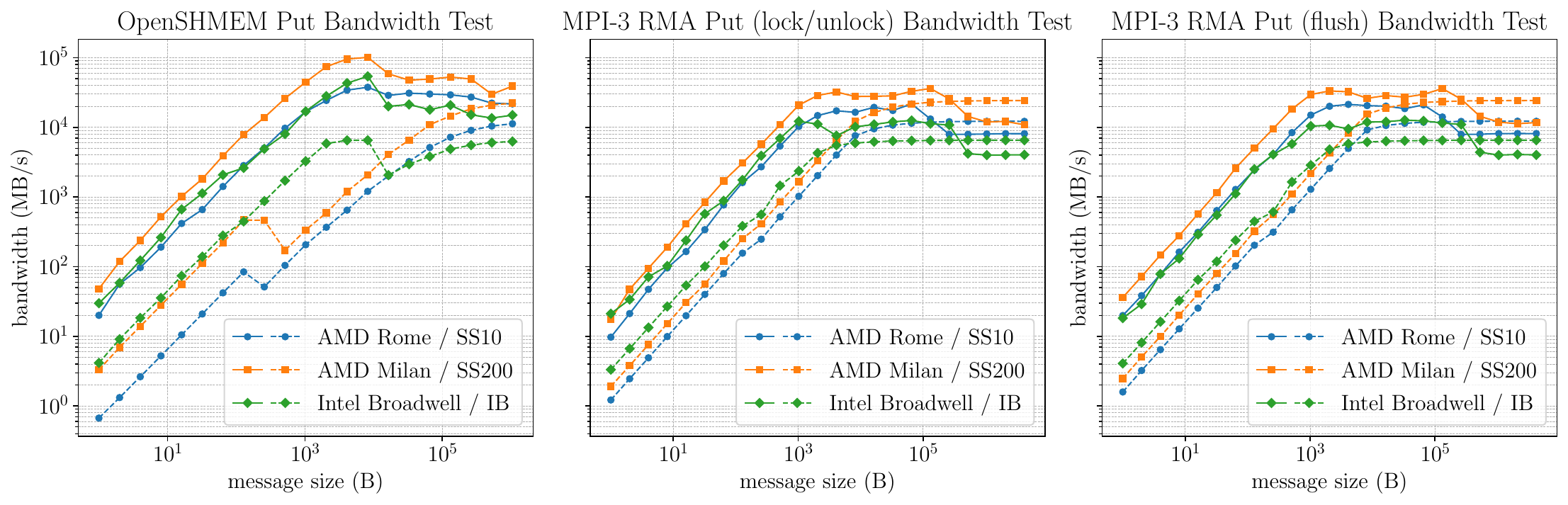}
    \caption{On-node (solid lines) and off-node (dashed lines) data transfer results of the OSU micro benchmark testing the bandwidth of remote put operations. The inter-node networks are HPE Slingshot 10 (SS10), HPE Slingshot 200 (SS200) and InfiniBand (IB).}
    \label{fig:osu_bandwidth_put}
\end{figure}

\begin{figure}[!htp]
    \centering
    \includegraphics[width=1.0\textwidth]{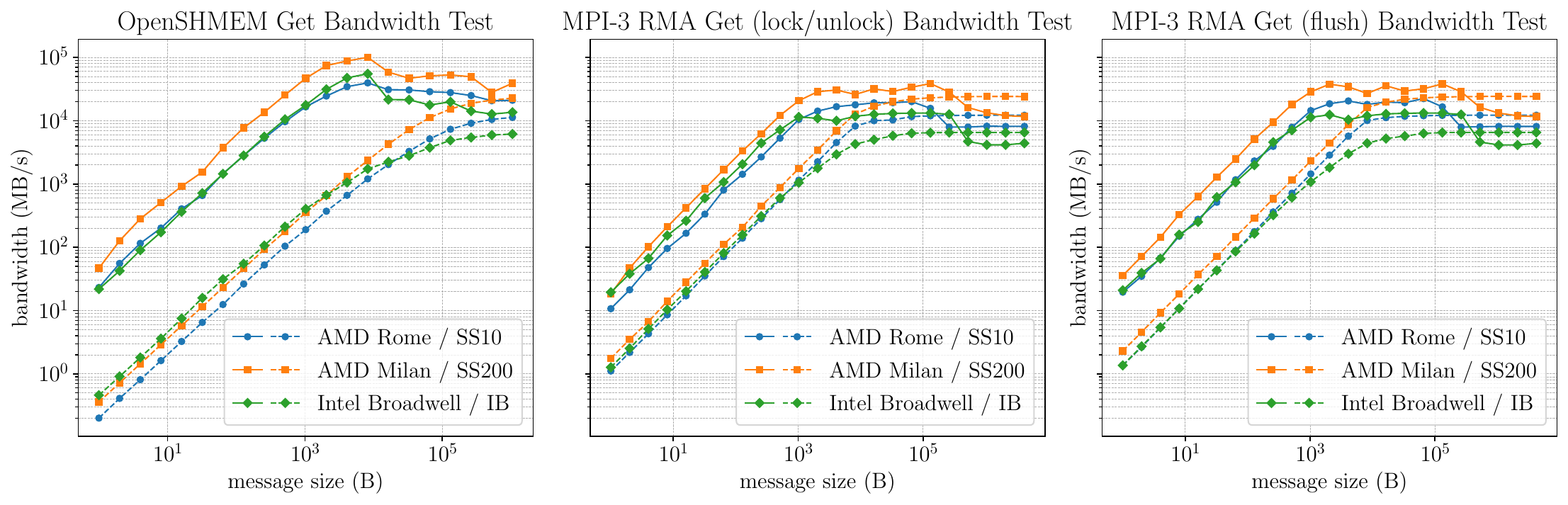}
    \caption{On-node (solid lines) and off-node (dashed lines) data transfer results of the OSU micro benchmark testing the bandwidth of remote get operations. The inter-node networks are HPE Slingshot 10 (SS10), HPE Slingshot 200 (SS200) and InfiniBand (IB).}
    \label{fig:osu_bandwidth_get}
\end{figure}

In order to find the nearest neighbour, the parcel cluster algorithm performs several MPI point-to-point communications between processes sharing computation domain boundaries. The latency and bandwidth of MPI point-to-point on the different computing systems is shown in \cref{fig:osu_mpi_p2p}. The latencies and bandwidths of the different computing systems are very close to each other, especially at lower message size. As expected, the latency is smaller for the intra-node communication. Similarly, the bandwidth is higher compared to off-node data transfers. However, their bandwidths are relatively close to each other. The left panel of \cref{fig:osu_mpi_p2p} shows the latency of MPI allreduce using
all physical cores available on a single compute node, i.e. 36 cores on Cirrus, 128 cores on ARCHER2 and 128 cores on Hotlum. The latency of the off-node data transfer is evaluated using the same number of cores like the intra-node MPI allreduce, however, distributed over two compute nodes such that each node has its half of the processes residing on a single CPU. The latencies of the intra-node and inter-node data transfers are very close, also among the different computing systems.
Next, we analyse the bandwidths and latencies of the one-sided communication routines.
For remote put operations with message sizes \SIrange{10}{100}{KB}, IB exhibits the lowest latency as shown in \cref{fig:osu_latency_put}. SS10 and SS200 demonstrate similar latencies over the entire measuring range. The SHMEM routine generally performs best, especially for message sizes beyond \SI{100}{KB}. In the case of intra-node communication, AMD Milan performs best apart for MPI-3 RMA with lock/unlock synchronisation up to about \SI{10}{KB}. A very similar behaviour is observed for the remote get operation as shown in \cref{fig:osu_latency_get}. As expected, the latency for intra-node communication is faster compared to inter-node communication across the network.
The bandwidth measurements illustrated in \cref{fig:osu_bandwidth_put} and \cref{fig:osu_bandwidth_get} for the remote put and remote get operation, respectively, clearly show the superiority of inter-node communication over the SS200 network. The bandwidth is for most of the message sizes twice as high as with SS10. IB and SS10 perform similarly for message sizes less than \SI{1}{KB}. The bandwidths using the shared memory within a NUMA region are generally closer among all computing systems, apart for the SHMEM put routine.\\
During the DG pruning step, the parcel cluster algorithm communicates single Fortran logicals (4 bytes) between processes. We therefore summarised the bandwidths and latencies for on-node and off-node 4-byte data transfers in \cref{tab:bw4byte} and \cref{tab:latency4byte}, respectively. While IB achieves the best bandwidth among all off-node transfers, SS10 performs worst. We therefore expect the parcel cluster algorithm to scale the least on ARCHER2.
\begin{table}[!htp]
    \small\sf\centering
    \caption{Bandwidth (\si{MB/s}) of 4-byte data transfers for on-node and off-node communication. In the case of MPI-3 RMA we specify the bandwidths of the lock/unlock tests.}
    \label{tab:bw4byte}
    \begin{center}
    \begin{tabular}{l
                    l
                    l
                    S[table-format=3.2]
                    S[table-format=3.2]
                    S[table-format=3.2]
                    S[table-format=3.2]
                    S[table-format=3.2]}
        \toprule
        {system} & & {data transfer}
                   & {MPI P2P}
                   & \multicolumn{2}{c}{MPI-3 RMA}
                   & \multicolumn{2}{c}{SHMEM} \\
        \cmidrule(rl){5-6}
        \cmidrule(rl){7-8}
                 & & & & {put} & {get} & {put} & {get} \\

        \midrule
        \multirow{2}{*}{ARCHER2}
            & AMD Rome & (on-node) & 18.75 & 47.21 & 47.86 & 96.85 & 115.27 \\
            & SS10 & (off-node)    &  7.26 &  4.91 &  4.30 &  2.63 &   0.81 \\
        \midrule
        \multirow{2}{*}{Hotlum}
            & AMD Milan & (on-node) & 39.77 & 94.90 & 101.81 & 238.10 & 281.69 \\
            & SS200 & (off-node)     &  9.37 &  7.58 &   5.13 &  13.92 &   1.44 \\
        \midrule
        \multirow{2}{*}{Cirrus}
            & Intel Broadwell & (on-node) & 34.63 & 71.26 & 66.80 & 122.70 & 89.89 \\
            & IB & (off-node)             & 14.21 & 13.36 &  6.68 &  18.37 &  1.81 \\
        \bottomrule
    \end{tabular}
    \end{center}
\end{table}
\begin{table}[!htp]
    \small\sf\centering
    \caption{Latency (\si{\micro s}) of 4-byte data transfers for on-node and off-node communication. In the case of MPI-3 RMA we specify the latencies of the lock/unlock tests.}
    \label{tab:latency4byte}
    \begin{center}
    \begin{tabular}{l
                    l
                    l
                    S[table-format=3.2]
                    S[table-format=3.2]
                    S[table-format=3.2]
                    S[table-format=3.2]
                    S[table-format=3.2]}
        \toprule
        {system} & & {data transfer}
                   & {MPI P2P}
                   & \multicolumn{2}{c}{MPI-3 RMA}
                   & \multicolumn{2}{c}{SHMEM} \\
        \cmidrule(rl){5-6}
        \cmidrule(rl){7-8}
                 & & & & {put} & {get} & {put} & {get} \\

        \midrule
        \multirow{2}{*}{ARCHER2}
            & AMD Rome & (on-node)  & 0.43 &  2.71 &  2.66 & 0.06 & 0.04 \\
            & SS10     & (off-node) & 2.48 & 18.97 & 15.64 & 6.34 & 4.97 \\
        \midrule
        \multirow{2}{*}{Hotlum}
            & AMD Milan & (on-node)  & 0.21 &  0.86 &  0.86 & 0.02 & 0.02 \\
            & SS200      & (off-node) & 1.72 & 18.64 & 18.21 & 5.59 & 2.74 \\
        \midrule
        \multirow{2}{*}{Cirrus}
            & Intel Broadwell & (on-node)  & 0.25 & 0.42 & 0.32 & 0.09 & 0.05 \\
            & IB              & (off-node) & 1.72 & 4.23 & 5.10 & 1.89 & 2.32 \\
        \bottomrule
    \end{tabular}
    \end{center}
\end{table}

\subsection{Parcel cluster algorithm}
In this section we report cluster (or merger) statistics as well as the parallel performance of the algorithm by means of two examples. The first example demonstrates the parallel scalability using artificial parcel configurations. The second example uses parcel configurations that are obtained from a fully evolved Rayleigh-Taylor flow instability simulation set up as in \citep{Frey2023}.

Before assessing the performance of the parallel algorithm, however, we verified its correctness. For this purpose, we compared the results obtained with $n/4$ ($n = 1,\hdots 8$) compute nodes on each machine to the serial version of the algorithm. The test domain spans the cube $[0, 1]^3$ and is discretised with $32^3$ isotropic grid cells. A parcel configuration is generated by randomly sampling $40$ parcels per grid cell where each parcel is assigned random parcel properties according to \cref{tab:sampling_distr}.
\begin{table}[!htp]
    \small\sf\centering
    \caption{Sampling parameters for artificial parcel configurations. Each parcel attribute is sampled from a uniform distribution $\unif(a, b)$ with lower bound $a$ and upper bound $b$.}
    \label{tab:sampling_distr}
    \begin{center}
    \begin{tabular}{ll}
        \toprule
        parcel attribute                  & distribution \\
        \midrule
         vorticity, $(\xi, \eta, \zeta)$  & $\unif(-10, 10)$ \\
         buoyancy, $b$                    & $\unif(-1, 1)$ \\
         volume, $V$                      & $\unif(V_{\min}/2, 3V_{min}/2)$ \\
         aspect ratio, $\lambda_1=a/c$    & $\unif(1, 4)$ \\
         aspect ratio, $\lambda_2=a/b$    & $\unif(1, 4)$ \\
         azimuthal angle, $\theta$        & $\unif(0, 2\pi)$ \\
         polar angle, $\phi$              & $\unif(0, \pi)$ \\
        \bottomrule
    \end{tabular}
    \end{center}
\end{table}
Their volumes are uniformly sampled between $[V_{\min}/2, 3V_{\min}/2]$ where $V_{\min} = V_{\mathrm{cell}} / 40$ denotes the parcel volume threshold below which a parcel is considered small and therefore marked for merging.

\subsubsection{Example: Artificial parcel configuration}\label{sec:example_art}
\begin{figure}[!htp]
    \centering
    \includegraphics[width=1.0\textwidth]{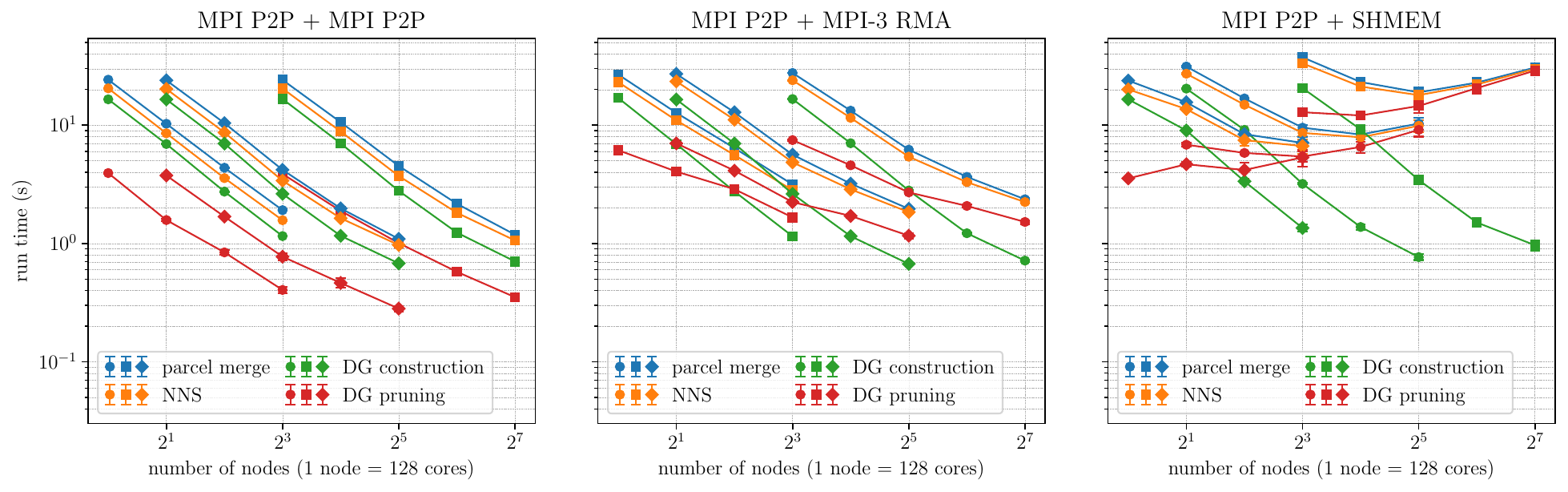}
    \caption{Combined parallel strong and weak scaling plot of the artificial parcel configuration example ran on the HPE Cray EX (ARCHER2) computing system with HPE Slingshot 10 interconnect and two AMD Rome processors per node. Each data point shows the maximum execution time across all MPI ranks averaged over 10 runs. The timing of the nearest neighbour search (NNS) (orange line) includes the timings of building the directed graphs (green line) and pruning the graphs (red line).}
    \label{fig:archer2-cray-random-scaling}
\end{figure}

\begin{figure}[!htp]
    \centering
    \includegraphics[width=1.0\textwidth]{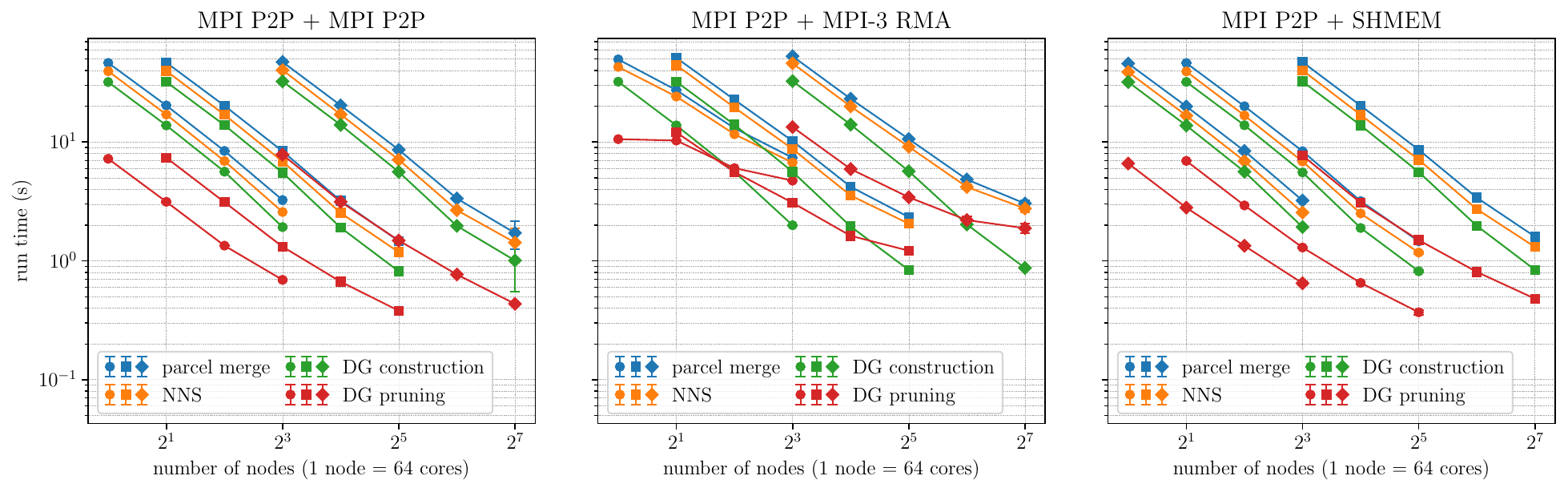}
    \caption{Combined parallel strong and weak scaling plot of the artificial parcel configuration example ran on the HPE Cray EX (Hotlum) computing system with HPE Slingshot 200 interconnect and two AMD Milan processors per node. Each data point shows the maximum execution time across all MPI ranks averaged over 10 runs. The timing of the nearest neighbour search (NNS) (orange line) includes the timings of building the directed graphs (green line) and pruning the graphs (red line).}
    \label{fig:hotlum-cray-random-scaling}
\end{figure}

\begin{figure}[!htp]
    \centering
    \includegraphics[width=\textwidth]{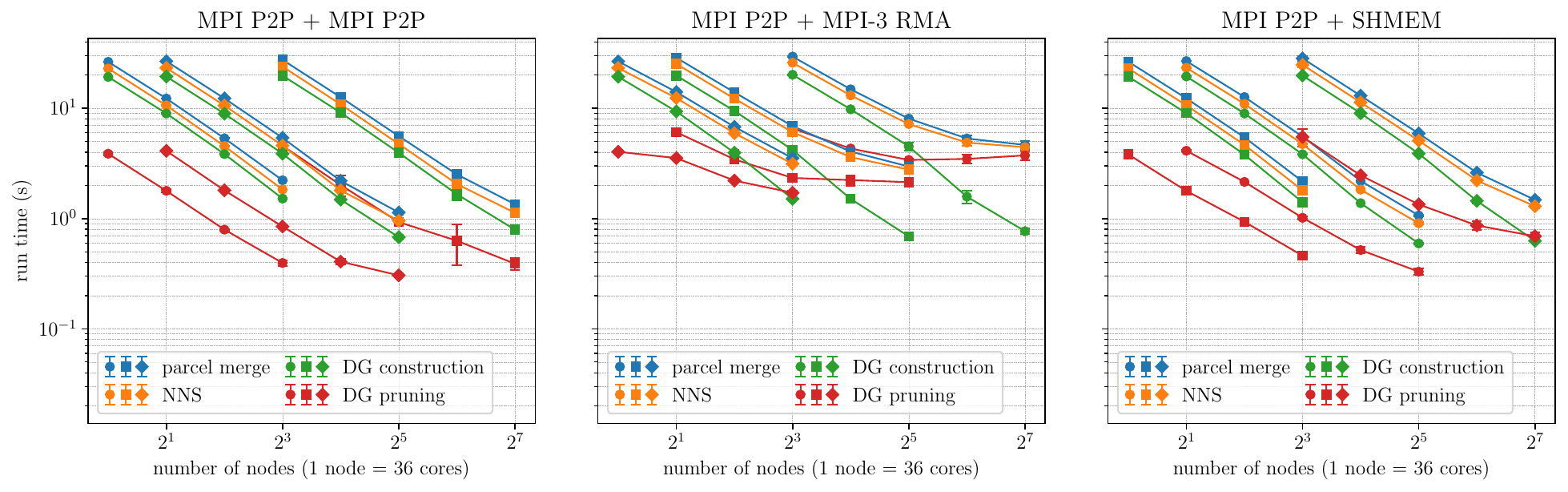}
    \caption{Combined parallel strong and weak scaling plot of the artificial parcel configuration example ran on the HPE SGI Apollo 8600 (Cirrus) computing system with InfiniBand interconnect and two Intel Broadwell processors per node. Each data point shows the maximum execution time across all MPI ranks averaged over 10  runs. The timing of the nearest neighbour search (NNS) (orange line) includes the timings of building the directed graphs (green line) and pruning the graphs (red line).}
    \label{fig:cirrus-gnu-random-scaling}
\end{figure}

The parallel strong and weak scaling on the different computing systems is shown in \cref{fig:archer2-cray-random-scaling,fig:hotlum-cray-random-scaling,fig:cirrus-gnu-random-scaling}. Each data point represents the average time over 10 runs where each run does 10 merge cycles, not including the cost of generating the artificial parcel configuration. Note that we always report the elapsed time of the slowest process. The standard deviation evaluated from the 10 runs is indicated by the error bars in the figure. All cycles start with a random sample of 20 parcels per grid cell where each parcel is assigned random properties as summarised in \cref{tab:sampling_distr}. The merge operation per cycle subsequently reduces the total number of parcels by about 36\%. The benchmarks are performed for an isotropic grid of mesh spacing $\Delta x = \Delta y = \Delta z = 5/16$. Due to the disparity of compute node configurations between HPE Cray EX and HPE SGI Apollo 8600, we varied the problem size to ensure similar workload. On ARCHER2 and Hotlum, we perform this scaling study with $256\times 512\times 32$, $512^2\times 32$ and $1024^2\times 32$ grid cells. On Cirrus, on the other hand, we carry out the benchmark with $144\times 288\times 32$,
$288^2\times 32$ and $576^2\times 32$ grid cells where we adapted the domain sizes in order to keep the isotropic grid mesh spacing at $5/16$. Note that we choose a different number of horizontal and vertical grid points for these tests as this is representative of our target applications (atmospheric case studies), and the 2D domain decomposition we have pursued. While on ARCHER2 and Cirrus we use all available physical cores per node, we only use half of the cores per node on Hotlum. The reason for this is because when we are running the MPI + SHMEM version of the code each of MPI and SHMEM requires a core. Therefore, the core limit per Network Interface Card (NIC) is limited to 126. As Hotlum has only one NIC per node this limits us to a maximum of 126 cores per node, we reduced this to 64 as it is the nearest power of two so we can more easily draw comparisons with ARCHER2.\\
In \cref{fig:archer2-cray-random-scaling,fig:hotlum-cray-random-scaling,fig:cirrus-gnu-random-scaling}, the timing denoted by \textit{parcel merge} (blue line) is the total time of the algorithm and therefore includes the timings of the nearest neighbour search (NNS) (orange line), directed graph (DG) construction (green line) and DG pruning (red line). Note that the total time excludes the time to set up the parcel configurations. On ARCHER2 (AMD Rome and SS10), the communication in the DG pruning step performed with MPI point-to-point (P2P) and MPI-3 RMA communication show good scaling behaviour. However, the code version calling SHMEM routines performs poorly. On Hotlum (AMD Milan and SS200), the scalings of MPI P2P and MPI-3 RMA are similar to the results on ARCHER2. However, the SHMEM version performs much better on Hotlum. Based on the bandwidth measurements (cf.\ \cref{tab:bw4byte}) of the OSU micro-benchmarks in \cref{sec:osu_benchmarks}, these results are not surprising. To obtain a single data point for \cref{fig:archer2-cray-random-scaling,fig:hotlum-cray-random-scaling,fig:cirrus-gnu-random-scaling}, the DG pruning algorithm performs about 43k to 247k remote put operations of 4-byte messages (Fortran logicals) depending on the number of executing processes and the problem size. Similarly, the number of remote get operations is in the order of 20k to 118k. The number of floating point operations per byte in memory is negligibly small and therefore the nearest neighbour search and especially the DG pruning step are memory- and communication-bound. Due to the vast amount of 4-byte memory data transfers between processes, the performance of the algorithm heavily depends on the latency and bandwidth for off-node communication. Since the on-node data transfer is generally higher than off-node data transfer (cf.\ \cref{tab:bw4byte}), it is mainly the interconnect that represents the performance limiting factor. The OSU micro-benchmarks have demonstrated that the off-node bandwidth of SHMEM put and SHMEM get for 4-byte messages on ARCHER2 are approximately 5.29 times and 1.78 times, respectively, lower compared to them on Hotlum. However, the number of remote put operations is at least double the amount of remote get operations, hence the performance should be mainly dominated by remote put. Since the MPI-3 RMA version of the DG pruning algorithm performs well on SS10, the bandwidth threshold to obtain a decent scaling behaviour must therefore be around 4.3 MB/s (cf.\ \cref{tab:bw4byte}).



\subsubsection{Example: Rayleigh-Taylor instability}\label{sec:example_rt}
\begin{figure}[!htp]
    \centering
    \includegraphics[width=1.0\textwidth]{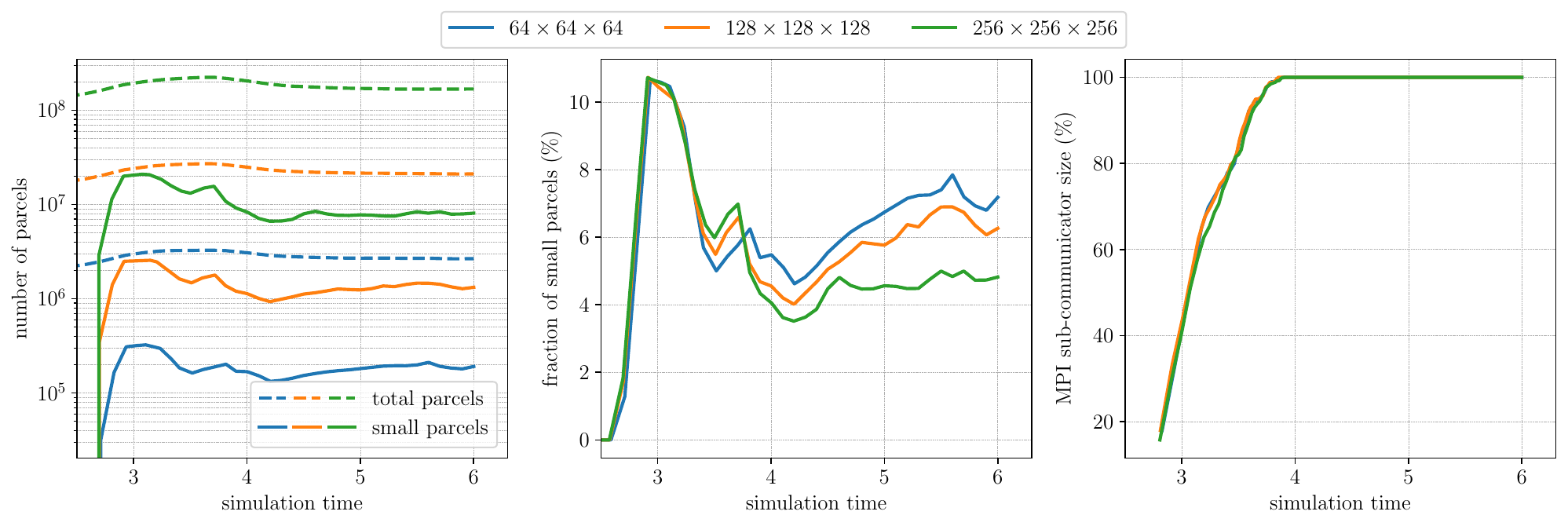}
    \caption{Evolution of the number of parcels (left panel), percentage of small parcels (centre panel) and the MPI sub-communicator size of the cluster algorithm (right panel). The first invocation of the cluster algorithm is around time $t \approx 2.8$. Note that we truncated the horizontal axis at simulation time 2.5. Our benchmark tests are performed with parcel configurations generated around time $t=3$ and $t=6.$}
    \label{fig:rt-subcomm}
\end{figure}

\begin{figure}[!htp]
    \centering
    \includegraphics[width=0.66\textwidth]{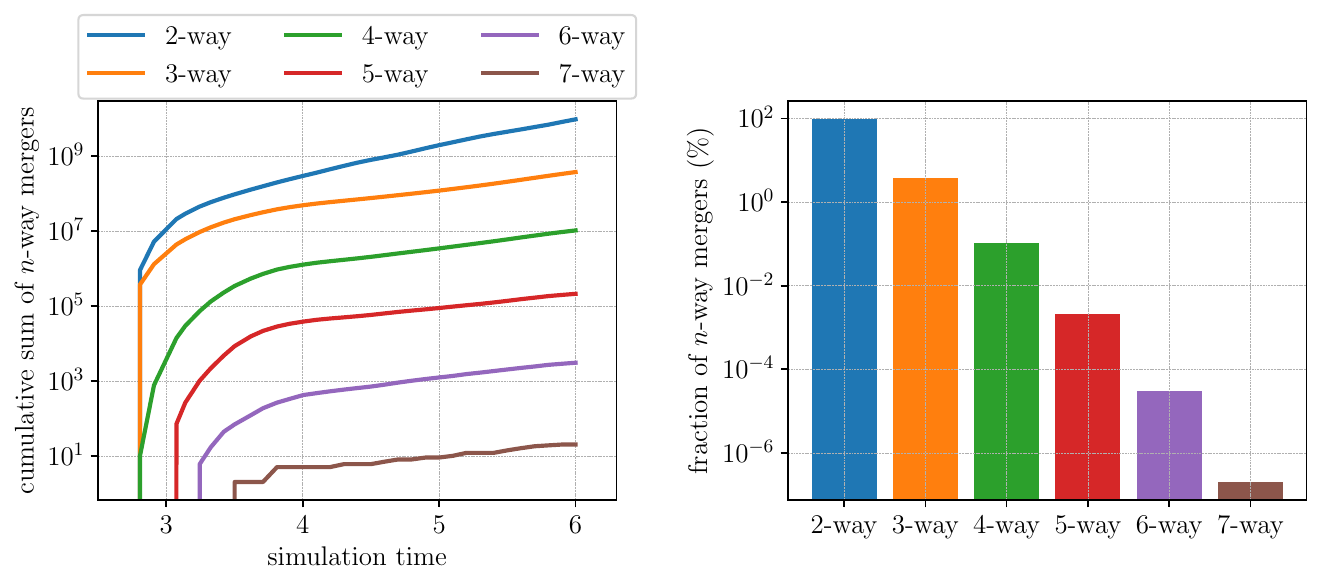}
    \caption{Cumulative sum of $n$-way cluster events (left panel) and percentage of total number of $n$-way cluster events (right panel) of the Rayleigh-Taylor instability test case up to time $t = 6$ for a grid resolution of $256^3$ grid cells.}
    \label{fig:rt-merger}
\end{figure}

\begin{figure}[!htp]
    \centering
    \includegraphics[width=0.66\textwidth]{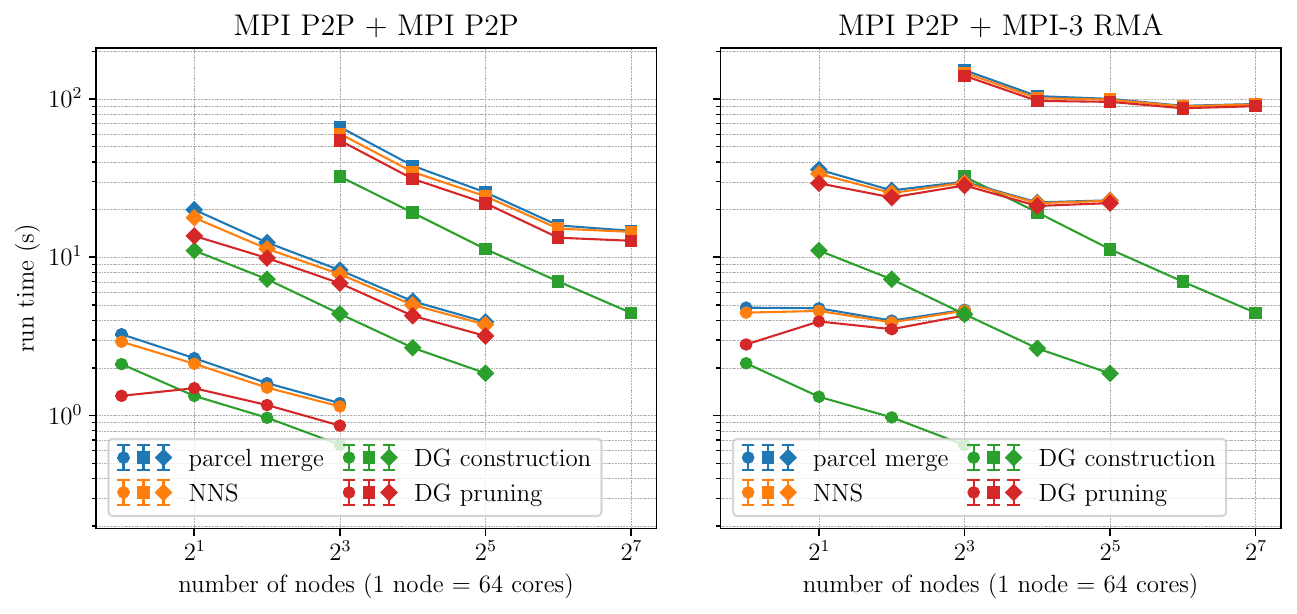}
    \caption{Parallel scaling plot of the Rayleigh-Taylor test case at early times $(t \approx 3)$ ran on the HPE Cray EX (ARCHER2) computing system with HPE Slingshot 10 interconnect and two AMD Rome processors per node. Each data point shows the maximum execution time across all MPI ranks averaged over 5 runs. The timing of the nearest neighbour search (NNS) (orange line) includes the timings of building the directed graphs (green line) and pruning the graphs (red line).}
    \label{fig:archer2-cray-read-early-scaling}
\end{figure}

\begin{figure}[!htp]
    \centering
    \includegraphics[width=1.0\textwidth]{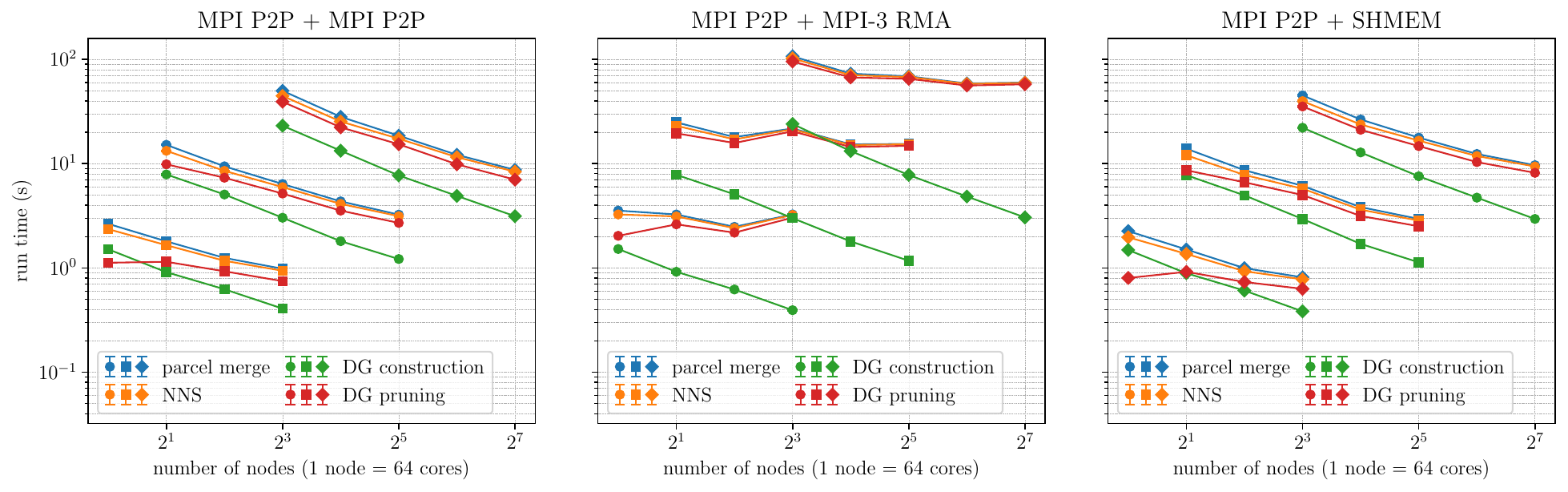}
    \caption{Parallel scaling plot of the Rayleigh-Taylor test case at early times $(t \approx 3)$ ran on the HPE Cray EX (Hotlum) computing system with HPE Slingshot 200 interconnect and two AMD Milan processors per node. Each data point shows the maximum execution time across all MPI ranks averaged over 5 runs. The timing of the nearest neighbour search (NNS) (orange line) includes the timings of building the directed graphs (green line) and pruning the graphs (red line).}
    \label{fig:hotlum-cray-read-early-scaling}
\end{figure}

\begin{figure}[!htp]
    \centering
    \includegraphics[width=1.0\textwidth]{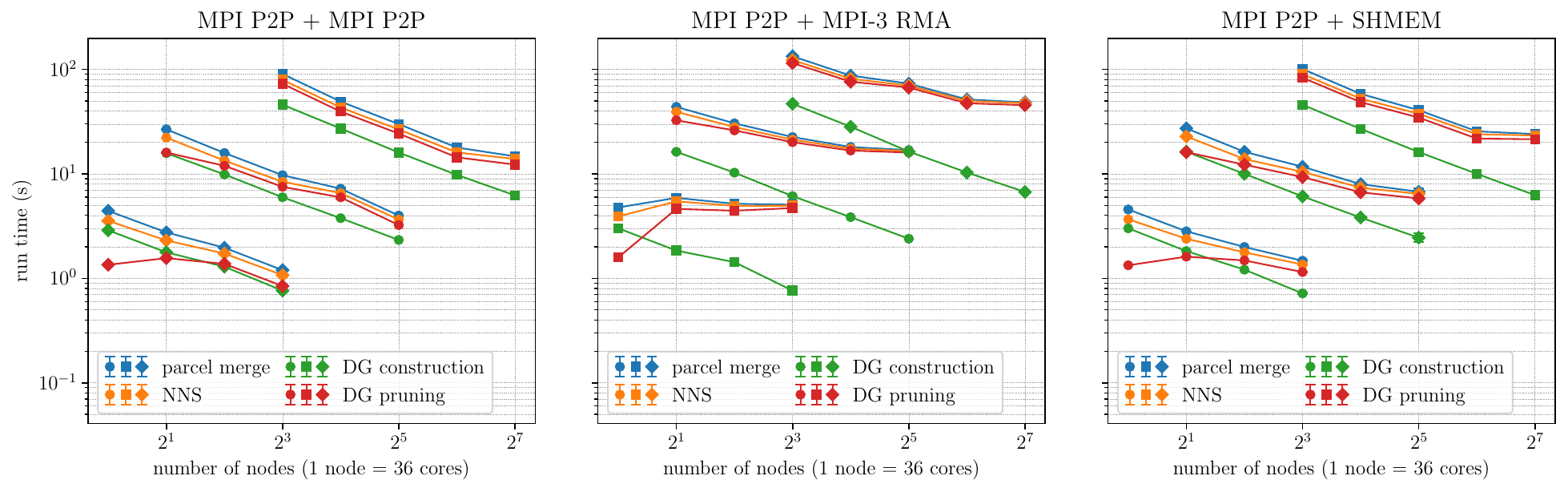}
    \caption{Parallel scaling plot of the Rayleigh-Taylor test case at early times $(t \approx 3)$ ran on the HPE SGI Apollo 8600 (Cirrus) computing system with InfiniBand interconnect and two Intel Broadwell processors per node. Each data point shows the maximum execution time across all MPI ranks averaged over 5 runs. The timing of the nearest neighbour search (NNS) (orange line) includes the timings of building the directed graphs (green line) and pruning the graphs (red line).}
    \label{fig:cirrus-gnu-read-early-scaling}
\end{figure}

\begin{figure}[!htp]
    \centering
    \includegraphics[width=0.66\textwidth]{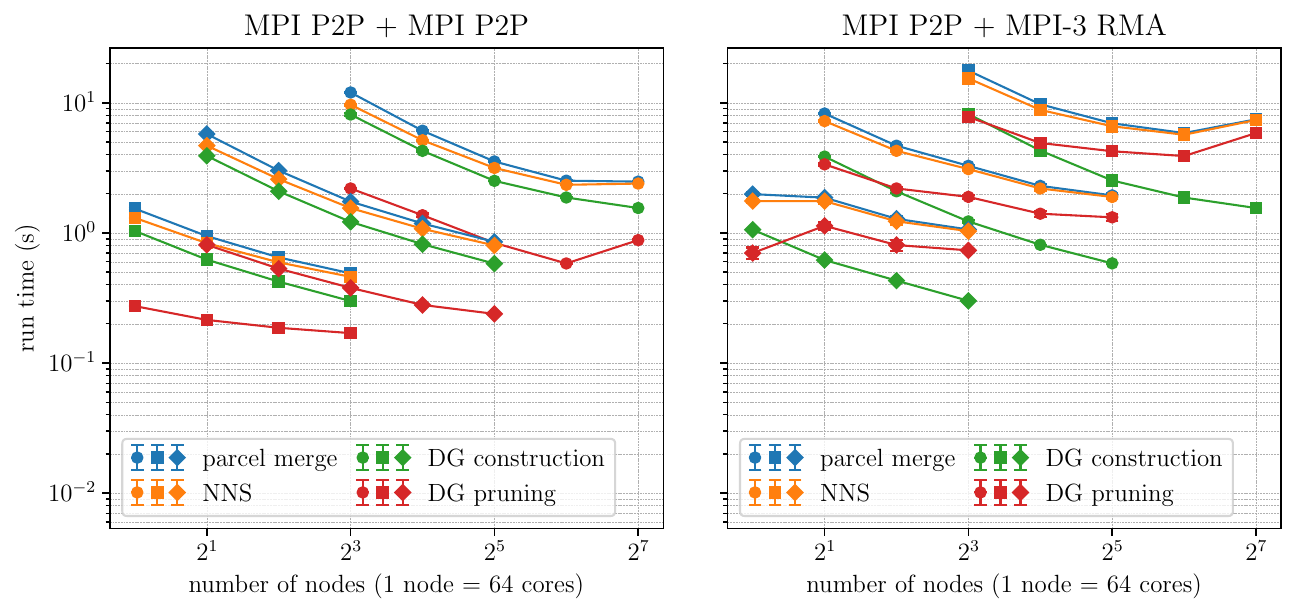}
    \caption{Parallel scaling plot of the Rayleigh-Taylor test case at late times $(t \approx 6)$ ran on the HPE Cray EX (ARCHER2) computing system with HPE Slingshot 10 interconnect and two AMD Rome processors per node. Each data point shows the maximum execution time across all MPI ranks averaged over 5 runs. The timing of the nearest neighbour search (NNS) (orange line) includes the timings of building the directed graphs (green line) and pruning the graphs (red line).}
    \label{fig:archer2-cray-read-late-scaling}
\end{figure}

\begin{figure}[!htp]
    \centering
    \includegraphics[width=1.0\textwidth]{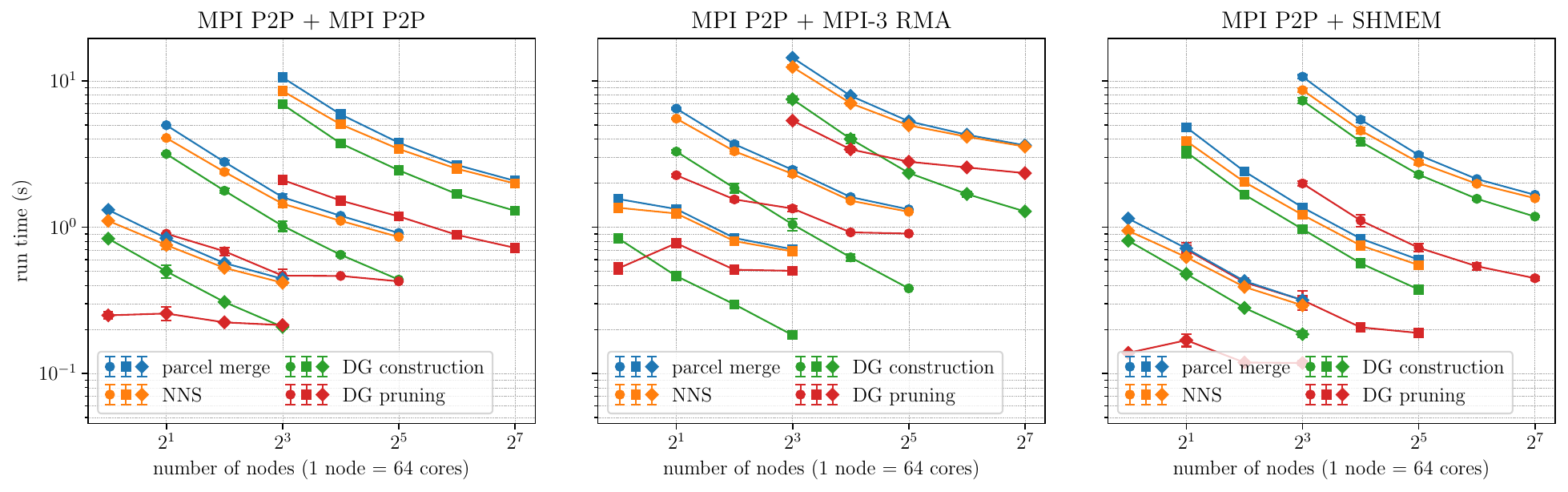}
    \caption{Parallel scaling plot of the Rayleigh-Taylor test case at late times $(t \approx 6)$ ran on the HPE Cray EX (Hotlum) computing system with HPE Slingshot 200 interconnect and two AMD Milan processors per node. Each data point shows the maximum execution time across all MPI ranks averaged over 5 runs. The timing of the nearest neighbour search (NNS) (orange line) includes the timings of building the directed graphs (green line) and pruning the graphs (red line).}
    \label{fig:hotlum-cray-read-late-scaling}
\end{figure}

\begin{figure}[!htp]
    \centering
    \includegraphics[width=1.0\textwidth]{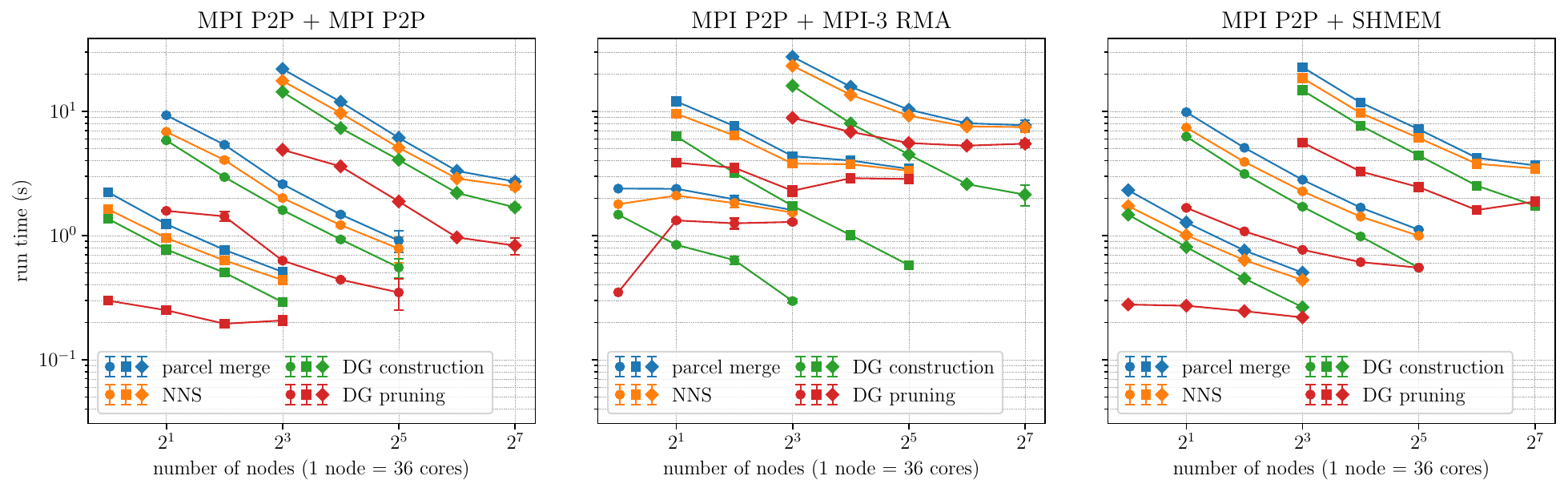}
    \caption{Parallel scaling plot of the Rayleigh-Taylor test case at late times $(t \approx 6)$ ran on the HPE SGI Apollo 8600 (Cirrus) computing system with InfiniBand interconnect and two Intel Broadwell processors per node. Each data point shows the maximum execution time across all MPI ranks averaged over 5 runs. The timing of the nearest neighbour search (NNS) (orange line) includes the timings of building the directed graphs (green line) and pruning the graphs (red line).}
    \label{fig:cirrus-gnu-read-late-scaling}
\end{figure}

In this benchmark we test the nearest neighbour cluster algorithm based on a real physical application, the Rayleigh-Taylor instability \citep{Rayleigh1882, Taylor1950}, where we use exactly the same setup as in \citep{Frey2023}, but with grid resolution consisting of $64^3$, $128^3$ and $256^3$ grid cells. The flow is triggered by an unstably stratified buoyancy profile that features a horizontal perturbation which causes the flow to overturn. In order to assess the parallel scalability we choose two stages of the flow that exhibit different  workload distributions. At early times the flow is inhomogeneous with respect to the occurrence of very small parcels resulting in an imbalanced workload during the cluster algorithm, visible in the right panel of \cref{fig:rt-subcomm}. The cluster algorithm is first invoked (the first parcels need merging) around time $t \approx 2.8$ with less than 20\% of processes. Note, we used 384, 512 and 768 processes to run the simulations with $64^3$, $128^3$ and $512$ grid cells.
At late times the flow is turbulent and the workload is homogeneously distributed.
As shown in the left panel of \cref{fig:rt-subcomm}, the total number of parcels in the simulation lies between $2\cdot 10^6$ and $2.24\cdot 10^8$ depending on the problem size. The fraction of small parcels (cf.\ the centre panel of \cref{fig:rt-subcomm}) remains always below 12\% throughout the simulated time. In \cref{fig:rt-merger} we report the cumulative sum of $n$-way mergers in the left panel and the total number of mergers in the right panel for the simulation with $256^3$ grid cells. As expected, 2-way mergers represent the majority of clusters with about 96\%, yet there are clusters involving up to 7 parcels.
For both stages of the simulation, we generated ten parcel configurations of subsequent time steps starting at either $t = 3$ or $t = 6$. In a bespoke benchmarking program we then iterate over the parcel configurations ten times (i.e. each parcel file is read in 10 times). The reported timings are averaged over five of these runs. Unfortunately, we cannot discuss SHMEM timings for the ARCHER2 system on this test case, because running these jobs would have been a waste of computing resources due to the poor scaling behaviour of SHMEM on SS10. Note that we only use 64 cores per node on ARCHER2 and Hotlum for the Rayleigh-Taylor benchmarks to increase the workload per core. \cref{fig:archer2-cray-read-early-scaling,fig:hotlum-cray-read-early-scaling,fig:cirrus-gnu-read-early-scaling} show the scaling at early simulation times ($3 < t < 3.5$). At this simulation stage the execution times of the cluster algorithm are dominated by the DG pruning (red line). A possible performance bottleneck of the DG pruning is process synchronisation. However, enabling the option to use a sub-communicator for MPI P2P and MPI-3 RMA (results not shown) to reduce this effect did not help improve the timings. On all computing systems we see poor scaling behaviour with MPI-3 RMA communication. However, the code versions with MPI P2P and SHMEM scale relatively well due to the higher inter-node bandwidth over the networks. At late times ($t> 6$), the DG pruning step is mainly hidden by the run time of the DG construction as observed in \cref{fig:archer2-cray-read-late-scaling,fig:hotlum-cray-read-late-scaling,fig:cirrus-gnu-read-late-scaling}. Since the number of small parcels is similar or even lower than at early times (cf.\ left panel of \cref{fig:rt-subcomm}), we suspect the more homogeneous workload distribution among the processes would slightly improve performance of MPI-3 RMA during the DG pruning.

\section{Conclusions}
\label{sec:conclusions}
We have presented the parallel performance of a cluster algorithm based on a nearest neighbour search using the Euclidean distance as a metric for proximity. The algorithm is used for the
merging of objects (here: parcels of ellipsoidal shape) in $N$-body simulations. The directed graphs (DGs) are simplified in an iterative procedure which can either make use of MPI point-to-point communication or one-sided communication based on MPI-3 RMA or SHMEM. We assessed the parallel performance on three supercomputers equipped with either HPE Slingshot 10, HPE Slingshot 200 or InifiniBand FDR interconnect. The benchmark test cases use artificial $N$-body configurations as well as configurations extracted from Rayleigh-Taylor fluid flow instability simulations. The artificial example demonstrates the parallel performance for a balanced workload scenario. Here, the parallel algorithm shows good strong and weak parallel scalability up to 16,384 processes on all three interconnects for grids ranging from $256\times 512\times 32$ to $1024^2\times 32$ grid cells on HPE Cray EX (ARCHER2 and Hotlum) and grids from $144\times 288\times 32$ to $576^2\times 32$ grid cells on HPE SGI Apollo 8600 (Cirrus). The benchmark clearly shows that SHMEM performs poorly over HPE Slingshot 10. For this test case, the cluster algorithm scales best with MPI P2P followed by MPI-3 RMA considering the performance over all networks. On the other hand, the parallel performance using parcel configurations generated from Rayleigh-Taylor instability simulations with $64^3$, $128^3$ and $256^3$ grid cells shows a different scaling result dependent on the workload distribution. At early times ($3 < t < 3.5$) the number of merge events is inhomogeneously distributed and thus results in an imbalanced workload and the timings are dominated by the DG pruning. Even though process synchronisation is required at several stages of the algorithm, it is not the bottleneck. At late times ($t > 6$), the flow exhibits homogeneous isotropic turbulence, thereby distributing the computational load more evenly. A detailed performance analysis based on the OSU micro-benchmarks has shown that the parallel scalability is limited by the bandwidth of the inter-node communication over the network during the DG pruning step. On HPE Slingshot 200 and InfiniBand, SHMEM outperforms MPI-3 RMA. However, we currently start and end an MPI-3 RMA access epoch for each individual remote put/get operation. In future work we may therefore explore the possibility to improve the performance of the MPI-3 RMA version of the code by initiating a single epoch for each stage of the DG pruning algorithm. We also plan to investigate other communication strategies.

\newpage
\subsection*{\normalsize\bfseries Code availability}
\noindent
The source code of the EPIC method (version 0.14.3) and of the parcel cluster algorithm are publicly available at \citep{Frey2025} and \url{https://github.com/EPIC-model/parcel-clustering}, respectively. The benchmark data and the associated plotting scripts are hosted on \url{https://github.com/EPIC-model/parcel-clustering-data}.

\subsection*{\normalsize\bfseries Acknowledgements}
We thank David Dritschel for his valuable feedback on the manuscript prior to submission. We further thank EPCC staff, especially David Henty and Linnea Gilius for their support on the Cirrus cluster. We would like to thank Hewlett Packard Enterprise for providing access to the HPE Cray EX Supercomputer, Hotlum. This work used the ARCHER2 UK National Supercomputing Service (\url{https://www.archer2.ac.uk}) which is funded and managed by UKRI and supported by EPCC, HPE and The University of Edinburgh. This work used the Cirrus UK National Tier-2 HPC Service at EPCC (\url{https://www.cirrus.ac.uk}) funded by the University of Edinburgh and EPSRC (EP/P020267/1). 

\subsection*{\normalsize\bfseries Funding}
The author(s) disclosed receipt of the following financial support for the research, authorship, and/or publication of this article: This work was supported by the UK Engineering and Physical Sciences Research Council [grant numbers EP/T025301/1, EP/T025409/1].

\bibliographystyle{apalike}

\end{document}